\documentclass[aps,twocolumn,prd,showpacs,showkeys,preprintnumbers,superscriptaddress,nofootinbib,floatfix,longbibliography]{revtex4-2}[prd]

\usepackage{orcidlink}
\usepackage{lipsum}
\usepackage{amsmath}
\usepackage{amsfonts}
\usepackage{amssymb}
\usepackage{mathrsfs}
\usepackage{graphicx}
\usepackage{color}
\usepackage{bm}
\usepackage{blindtext}
\usepackage{wasysym}
\usepackage{hyperref}
\hypersetup{colorlinks=true,allcolors=blue}
\usepackage{adjustbox}
\usepackage[normalem]{ulem}
\usepackage{fontawesome} 
\usepackage{lineno}
\usepackage{multirow}

\usepackage{amsmath}
\newcommand{\mc}[3]{\multicolumn{#1}{#2}{#3}}
\newcommand{\mr}[3]{\multirow{#1}{#2}{#3}}

\definecolor{mygreen}{RGB}{0,120,0}

\begin{document}

\title{Constraints on long-range neutrino interactions from a variety of $U(1)^\prime$ symmetries using atmospheric neutrinos at IceCube DeepCore}

\author{Gopal Garg\,\orcidlink{0009-0003-3134-2089}}
\affiliation{Institute of Physics, Sachivalaya Marg, Sainik School Post, Bhubaneswar 751005, India}
\affiliation{Department of Physics, Aligarh Muslim University, Aligarh 202002, India}
	
\author{J Krishnamoorthi\,\orcidlink{0009-0006-1352-2248}}
\affiliation{Institute of Physics, Sachivalaya Marg, Sainik School Post, Bhubaneswar 751005, India}
\affiliation{Department of Physics, Aligarh Muslim University, Aligarh 202002, India}

\author{Anil Kumar\,\orcidlink{0000-0002-8367-8401}}
\affiliation{Institute of Physics, Sachivalaya Marg, Sainik School Post, Bhubaneswar 751005, India}

\author{Sanjib Kumar Agarwalla\,\orcidlink{0000-0002-9714-8866}}
\affiliation{Institute of Physics, Sachivalaya Marg, Sainik School Post, Bhubaneswar 751005, India}
\affiliation{Homi Bhabha National Institute, Training School Complex, Anushakti Nagar, Mumbai 400094, India\\
    {\tt gi8820@myamu.ac.in, krishnamoorthi.j@iopb.res.in, anil.k@iopb.res.in, sanjib@iopb.res.in} \smallskip}

\preprint{IOP/BBSR/2026-07}

\date{\today}

\begin{abstract}
	Neutrino oscillation experiments provide a unique probe to search for the physics beyond the Standard Model. In this work, we search for a broad class of anomaly-free flavor-dependent $U(1)^\prime$ symmetries using atmospheric neutrino data for the first time. Gauging these $U(1)^\prime$ symmetries give rise to ultra-light vector gauge bosons mediating long-range interactions (LRI) of neutrinos. These new interactions are sourced by the matter present in local and distant Universe, which can affect oscillations of neutrinos passing through the Earth. We use 8 years of high-purity $\nu_\mu$ charged-current neutrino events from IceCube DeepCore to search for these new interactions. We find no evidence for such new interactions in the data sample and place stringent constraints on the corresponding LRI potentials. These results are also translated as the bounds on the coupling strength and mass of mediator over their wide ranges for a plethora of $U(1)^\prime$ symmetries.
\end{abstract}


\maketitle

\section{Introduction and motivation}
\label{sec:intro}

The phenomenon of neutrino flavor oscillations~\cite{Super-Kamiokande:1998kpq,IceCube:2014flw} has emerged as one of the most sensitive tools for probing physics beyond the Standard Model (BSM). The neutrino flavor transitions arise from quantum interference among neutrino mass eigenstates and are further modified by matter effects during propagation of neutrinos. As a consequence, even extremely small new interactions can induce observable deviations in neutrino oscillation probabilities. This remarkable sensitivity makes neutrino oscillation experiments uniquely suited to explore new feebly interacting BSM physics scenarios.

Among the wide range of proposed extensions of the Standard Model (SM), a particularly well-motivated class involves new flavor-dependent neutrino interactions~\cite{He:1990pn, Foot:1990mn,Foot:1990uf,He:1991qd,Foot:1994vd,Lee:1955vk,Okun:1995dn,Dolgov:1999gk,Langacker:2008yv}. In such frameworks, neutrinos experience a new additional force, beyond the standard weak interactions, with couplings that are not universal across the three flavors, $\nu_e$, $\nu_\mu$, and $\nu_\tau$. Such flavor-dependent interactions introduce additional terms in the effective Hamiltonian governing neutrino propagation, thereby modifying the three-flavor neutrino oscillation probabilities in a characteristic way that can be tested experimentally~\cite{Joshipura:2003jh,Bandyopadhyay:2006uh,Bustamante:2018mzu,Coloma:2020gfv,Agarwalla:2023sng,Garg:2026gwx,Grifols:2003gy,Heeck:2010pg,Davoudiasl:2011sz,Singh:2023nek,Agarwalla:2024ylc}.

These new interactions are expected to be significantly weaker than the standard weak interactions. However, if they are mediated by an ultra-light gauge boson, they can give rise to long-range interactions (LRI). In such scenarios, neutrinos propagating through the Earth experience not only the local matter potential generated by the Earth itself, but also additional contributions from the Moon, Sun, Milky Way, and cosmological matter distribution~\cite{Bustamante:2018mzu}. The cumulative contribution from these massive sources can significantly enhance the impact of otherwise feeble interactions, leading to observable modifications in neutrino flavor transitions. Consequently, this makes neutrino oscillation experiments a uniquely powerful avenue for probing the signatures of a BSM physics scenarios associated with a new fifth force mediated by an ultra-light gauge boson in the neutrino sector.

Such interactions arise naturally in the extensions of the Standard Model featuring additional $U(1)^\prime$ gauge symmetries. These symmetries can be constructed by gauging anomaly-free combinations of the individual lepton numbers ($L_e$, $L_\mu$, and $L_\tau$) as well as the total Lepton ($L$) and baryon numbers ($B$); see Refs.~\cite{He:1990pn, Foot:1990mn, He:1991qd,Foot:1994vd,Langacker:2008yv} for a comprehensive review. Promoting one of these $U(1)^\prime$ symmetries to a local gauge symmetry introduces a new neutral vector gauge boson, $Z^\prime$, which mediates Yukawa-like interaction between neutrinos and ordinary matter - namely electron, proton, and neutron (up and down quarks).

Since different lepton flavors have distinct $U(1)^\prime$ charges for the underlying symmetry, the induced interactions are flavor non-universal. Consequently, the resulting matter potentials affect $\nu_e$, $\nu_\mu$, and $\nu_\tau$ differently, thereby modifying their flavor evolution during propagation. The range of the new interaction is determined by the inverse of the mediator mass, with heavier mediators leading to effective point-like interactions~\cite{Wolfenstein:1977ue,Valle:1987gv,Guzzo:1991hi,Roulet:1991sm,Ohlsson:2012kf,Agarwalla:2012wf,Agarwalla:2014bsa,Agarwalla:2015cta,Farzan:2017xzy,Bhupal:2019qno,Kumar:2021lrn,Coloma:2023ixt,Krishnamoorthi:2025efw,Krishnamoorthi:2026zve}. On the other hand, lighter mediators correspond to the long-range interactions. In this work, we focus on ultra-light mediators with masses in the range from $10^{-35}\,\mathrm{eV}$ to $10^{-10}\,\mathrm{eV}$ corresponding to interaction lengths spanning from cosmological scales down to terrestrial distances. To date, there is no experimental evidence for the existance of LRI. However, a broad range of observations place stringent constraints on the viable LRI parameter space. Significant bounds on LRI parameters have been derived from neutrino oscillation studies~\cite{Joshipura:2003jh,Garg:2026gwx,Grifols:2003gy,Bandyopadhyay:2006uh,Gonzalez-Garcia:2006vic,Heeck:2010pg,Bustamante:2018mzu,Agarwalla:2023sng,Singh:2023nek,Agarwalla:2024ylc,Davoudiasl:2011sz,Wise:2018rnb,Dror:2020fbh,Coloma:2020gfv,Alonso-Alvarez:2023tii}, as well as from several probes other than neutrinos~\cite{Adelberger:2009zz, Salumbides:2013dua,Schlamminger:2007ht,Baryakhtar:2017ngi,KumarPoddar:2019ceq,KumarPoddar:2020kdz}. A concise overview of the existing bounds can be found in Ref.~\cite{Wise:2018rnb,Agarwalla:2024ylc}.

Atmospheric neutrinos span a broad range of energies from a few hundreds of MeV to more than TeV and propagation distances ranging from tens of kilometers to the diameter of Earth~\cite{Honda:2015fha}. This extensive coverage enhances their sensitivity to subleading effects, such as additional matter potentials that alter neutrino flavor evolution during their propagation. The IceCube DeepCore detector at the South Pole is sensitive to the GeV-energy atmospheric neutrinos~\cite{IceCube:2011ucd}. With its large instrumented volume and multi-year exposure, DeepCore has been established as a leading experiment for precision measurements of the atmospheric neutrino oscillation parameters as well as for the BSM physics searches~\cite{IceCube:2017lak,IceCubeCollaboration:2023wtb,IceCubeCollaboration:2024ssx,IceCube:2019dqi,IceCube:2017ivd,IceCube:2020phf,IceCube:2020tka,IceCubeCollaboration:2024nle,IceCube:2024pky,IceCube:2024dlz,IceCube:2017zcu,IceCubeCollaboration:2021euf,IceCube:2025kve}. This motivates us to use atmospheric neutrino data collected by DeepCore for probing subtle signatures of long-range interactions.

This analysis brings together several key ingredients - each previously explored in different contexts - into a unified framework. We evaluate the LRI potential generated by the matter distribution in local and distant Universe. We explore a wide class of flavor-dependent $U(1)^\prime$ symmetries (see Table~\ref{tab:symm}), each characterized by a distinct structure of the induced matter potential, and consequently, by different modifications to neutrino oscillation probabilities. While most of the previous analyses have focused on a limited subset of symmetries, primarily those sourced by electrons, our approach incorporates contributions from all relevant matter constituents - electrons, protons, and neutrons. We utilize the 8 years of high-purity $\nu_\mu$ charged-current (CC) events collected by IceCube DeepCore~\cite{DVN/B4RITM_2025} to place stringent constraints on the LRI parameters in the context of $U(1)^\prime$ symmetries. Our analysis incorporates a rigorous treatment of all the systematic uncertainties provided by the IceCube collaboration, which ensures the robustness of our results.

\begin{figure}
	\centering
	\includegraphics[width=\linewidth]{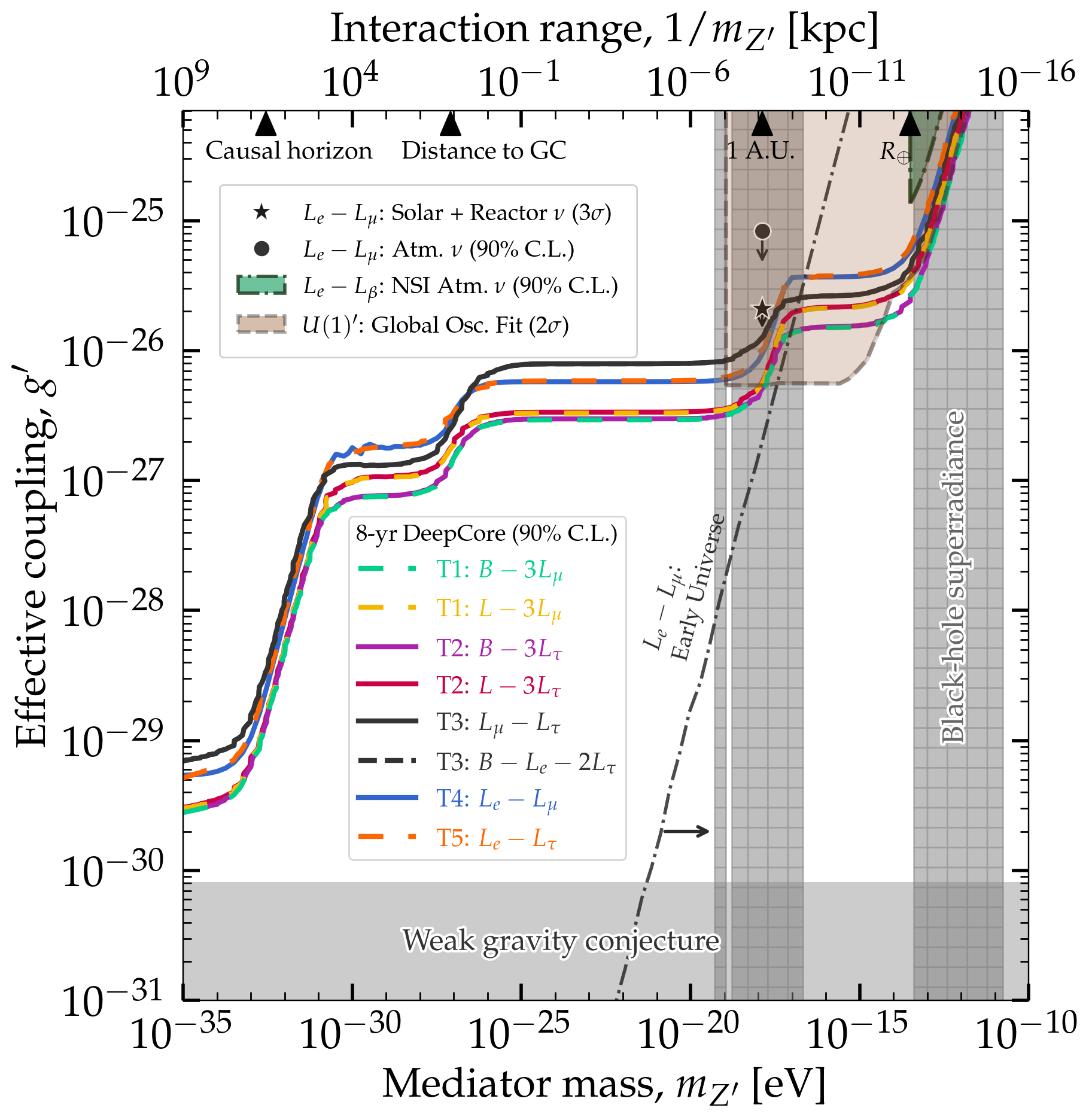}
	\caption{Constraints on the long-range neutrino interactions at 90\% C.L. in the plane of the effective coupling strength $g^\prime$ and the mediator mass $m_{Z^\prime}$ using the 8-year high-purity $\nu_\mu$ CC data of IceCube DeepCore. The colored curves, as shown in the legend box, correspond to the limits obtained for the different LRI textures, T1, T2, and T3, assuming normal mass ordering. These constraints on $g^\prime$ and $m_{Z^\prime}$ are obtained by translating the bounds on the LRI potential (see Eq.~\ref{eq:total_pot}) following the methodology outlined in Refs.~\cite{Bustamante:2018mzu,Agarwalla:2023sng,Agarwalla:2024ylc}. We have also shown the existing bounds from other analysis for comparison (see text for details). The arrows indicate the allowed parameter space, while the shaded area denote the excluded regions.}
	\label{fig:g_m_bound}
\end{figure}

We present the key results of this work in Fig.~\ref{fig:g_m_bound}, which shows the constraints on LRI coupling strength ($g^\prime$) and mass of mediating ultra-light gauge boson ($m_{Z^\prime}$) for a subset of $U(1)^\prime$ symmetries listed in first three rows of table~\ref{tab:symm} (T1, T2, and T3). In Fig.~\ref{fig:g_m_bound}, we also include the corresponding bounds for the symmetries $L_e - L_\mu$ and $L_e - L_\tau$ (T4 and T5 in table~\ref{tab:symm}) using same dataset from our previous study~\cite{Garg:2026gwx} for completeness. We do not consider T6 in this analysis due to its negligible effect on the $\nu_\mu$ CC data set. A more detailed discussion on these $U(1)^\prime$ symmetries is given in section~\ref{sec:formalism}.

The different colored isocontours in Fig.~\ref{fig:g_m_bound} represent the final constraints on the effective LRI potentials for various textures. A prominent feature of the figure is a characteristic step-like behavior as a function of the mediator mass, $m_{Z^\prime}$. As the interaction range increases with decrease in $m_{Z^\prime}$, more matter content starts to contribute, and the same LRI potential can be generated with smaller values of the coupling strength $g^\prime$ (see Eqs.~\ref{eq:V_Z} and~\ref{eq:V_ZZ'}). As we go towards left side in Fig.~\ref{fig:g_m_bound}, the step-like feature appears at a given mediator mass whenever the corresponding interaction range includes an additional matter object. For mediator masses in the range from $10^{-10}$ eV to $10^{-18}$ eV, the dominant contribution arises from the Earth and Moon. As the mediator mass decreases below $10^{-18}$ eV, the potential starts receiving the contribution from the Sun. At masses below $10^{-27}$ eV, matter in the Milky Way begins contributing to the potential. Finally, for mediator masses lower than $10^{-33}$ eV, the dominant contribution originates from the cosmological matter distribution.

In Fig.~\ref{fig:g_m_bound}, we also present the existing bounds from other neutrino-based analyses, which include atmospheric neutrinos~\cite{Joshipura:2003jh, Garg:2026gwx}, solar and reactor neutrinos~\cite{Bandyopadhyay:2006uh}, recent global oscillation fit~\cite{Coloma:2020gfv}, and searches for non-standard interactions~\cite{Super-Kamiokande:2011dam,Ohlsson:2012kf,Gonzalez-Garcia:2013usa}. In addition, we also show the complementary bounds from non-neutrino studies such as the early
Universe~\cite{Dror:2020fbh}, black-hole superradiance~\cite{Baryakhtar:2017ngi}, and the weak
gravity conjecture~\cite{Arkani-Hamed:2006emk}. The allowed region in the parameter space is shown by the direction of the arrows. Overall, our analysis using the IceCube DeepCore data improves the existing constraints over a broad range of LRI coupling strengths and ultra-light mediator mass for a wide class of $U(1)^\prime$ symmetries within a unified framework.

The paper is organized as follows. In section~\ref{sec:formalism}, we discuss the  $U(1)^\prime$ symmetries and LRI framework. Section~\ref{sec:Hamiltonian} discusses how LRI modify neutrino oscillations. Section~\ref{sec:DeepCore} describes the IceCube DeepCore detector. In section~\ref{sec:events}, we describe the data sample and the impact of LRI on the events distributions at IceCube DeepCore. Section~\ref{sec:results} presents our main results on LRI parameters. Finally, in section~\ref{sec:conclusion}, we summarize our findings and conclude. In appendix~\ref{app:U1-charges}, we provide the $U(1)^\prime$ charge assignments of fermions for the different symmetries. Appendix~\ref{app:Oscillograms} presents the neutrino oscillograms in the presence of LRI. We summarize the best-fit values of systematic uncertainty parameters in Appendix~\ref{app:systematics} and present the Data-MC agreement in Appendix~\ref{app:data-mc}.

\section{$U(1)^\prime$ symmetries and long-range neutrino interactions}
\label{sec:formalism}
The Standard Model possesses several accidental global $U(1)$ symmetries associated with the individual lepton numbers $L_e$, $L_\mu$, and $L_\tau$, as well as the total Lepton number $L$ and baryon number $B$. Promoting these symmetries individually to local gauge symmetries generally leads to anomalies. However, their specific linear combinations can be made anomaly-free, either within the SM particle content or by extending it to include right-handed neutrinos~\cite{Ma:1997nq,Araki:2012ip,Allanach:2018vjg}. These anomaly-free combinations provide well-motivated $U(1)^\prime$ extensions of the SM.

\begin{table*}[t!]
	\centering
	\footnotesize
	\begin{adjustbox}{width=\textwidth}
		\renewcommand{\arraystretch}{1.4}
		\begin{tabular}{|c|c|c|c|c|c|}
			\hline
			\multirow{3}{*}{\shortstack{Texture \\ of $\mathbf{V}_{\rm LRI}$}} &
			\multirow{3}{*}{\shortstack{$\mathbf{V}_{\rm LRI}$ matrix \\ structure}} &
			\multirow{3}{*}{$U(1)^\prime$ symmetry} &
			\multicolumn{3}{c|}{New matter potential, $\mathbf{V}_{\rm LRI} = \textrm{diag}(V_{{\rm LRI},e},\, V_{\rm LRI,\mu},\, V_{\rm LRI,\tau})$}
			\\ \cline{4-6}
			&
			&
			&
			\multirow{2}{*}{\shortstack{Texture to place limits, \\ $\mathbf{V}_{\rm LRI} = V_{\rm LRI} \cdot \textrm{diag}(\ldots)$}}
			&
			\multirow{2}{*}{\shortstack{General form \\ of $V_{\rm LRI}$, Eq.~\ref{eq:total_pot}}}
			&
			\multirow{2}{*}{\shortstack{Form of $V_{\rm LRI}$ to obtain limits \\ on $g^\prime$ vs.~$m_{Z^{\prime}}$, Eq.~\ref{eq:total_lri}}}
			\\
			& & & & &
			\\
			\hline
			\multirow{4}{*}{T1} &
			\multirow{4}{*}{$\left(\begin{array}{ccc} 0 & & \\ & \bullet & \\ & & 0 \end{array}\right)$} &
			\multirow{2}{*}{$B-3L_\mu$} &
			\multirow{2}{*}{$\textrm{diag}(0,-1,0)$} &
			\multirow{2}{*}{$3(V_p+V_n)$} &
			\multirow{2}{*}{$6(V^{\oplus}_{e}+V^{\leftmoon}_{e}+V^{\rm MW}_{e})+\frac{15}{4}V^{\astrosun}_{e}+\frac{24}{7}V^{\rm cos}_{e}$}
			\\
			& & & & &
			\\
			\cline{3-6}
			&
			&
			\multirow{2}{*}{$L-3L_\mu$} &
			\multirow{2}{*}{$\textrm{diag}(0,-1,0)$} &
			\multirow{2}{*}{$3V_e$} &
			\multirow{2}{*}{$3(V^{\oplus}_{e}+V^{\leftmoon}_{e}+V^{\rm MW}_{e}+V^{\astrosun}_{e}+V^{\rm cos}_{e})$}
			\\
			& & & & &
			\\
			\hline
			\multirow{4}{*}{T2} &
			\multirow{4}{*}{$\left(\begin{array}{ccc} 0 & & \\ & 0 & \\ & & \bullet \end{array}\right)$} &
			\multirow{2}{*}{$B-3L_\tau$} &
			\multirow{2}{*}{$\textrm{diag}(0,0,-1)$} &
			\multirow{2}{*}{$3(V_p+V_n)$} &
			\multirow{2}{*}{$6(V^{\oplus}_{e}+V^{\leftmoon}_{e}+V^{\rm MW}_{e})+\frac{15}{4}V^{\astrosun}_{e}+\frac{24}{7}V^{\rm cos}_{e}$}
			\\
			& & & & &
			\\
			\cline{3-6}
			&
			&
			\multirow{2}{*}{$L-3L_\tau$} &
			\multirow{2}{*}{$\textrm{diag}(0,0,-1)$} &
			\multirow{2}{*}{$3V_e$} &
			\multirow{2}{*}{$3(V^{\oplus}_{e}+V^{\leftmoon}_{e}+V^{\rm MW}_{e}+V^{\astrosun}_{e}+V^{\rm cos}_{e})$}
			\\
			& & & & &
			\\
			\hline
			\multirow{4}{*}{T3} &
			\multirow{4}{*}{$\left(\begin{array}{ccc} 0 & & \\ & \bullet & \\ & & \bullet \end{array}\right)$} &
			\multirow{2}{*}{$L_\mu-L_\tau$} &
			\multirow{2}{*}{$\textrm{diag}(0,1,-1)$} &
			\multirow{2}{*}{$-V_e+V_p+V_n$} &
			\multirow{2}{*}{$V^{\oplus}_{e}+V^{\leftmoon}_{e}+V^{\rm MW}_{e}+\frac{1}{4}V^{\astrosun}_{e}+\frac{1}{7}V^{\rm cos}_{e}$}
			\\
			& & & & &
			\\
			\cline{3-6}
			&
			&
			\multirow{2}{*}{$B-L_e-2L_\tau$} &
			\multirow{2}{*}{$\textrm{diag}(0,1,-1)$} &
			\multirow{2}{*}{$-V_e+V_p+V_n$} &
			\multirow{2}{*}{$V^{\oplus}_{e}+V^{\leftmoon}_{e}+V^{\rm MW}_{e}+\frac{1}{4}V^{\astrosun}_{e}+\frac{1}{7}V^{\rm cos}_{e}$}
			\\
			& & & & &
			\\
			\hline
			\multirow{3}{*}{T4} &
			\multirow{3}{*}{$\left(\begin{array}{ccc} \bullet & & \\ & \bullet & \\ & & 0 \end{array}\right)$} &
			\multirow{3}{*}{$L_e-L_\mu$} &
			\multirow{3}{*}{$\textrm{diag}(1,-1,0)$} &
			\multirow{3}{*}{$V_e$} &
			\multirow{3}{*}{$V^{\oplus}_{e}+V^{\leftmoon}_{e}+V^{\rm MW}_{e}+V^{\astrosun}_{e}+V^{\rm cos}_{e}$}
			\\
			& & & & &
			\\
			& & & & &
			\\
			\hline
			\multirow{3}{*}{T5} &
			\multirow{3}{*}{$\left(\begin{array}{ccc} \bullet & & \\ & 0 & \\ & & \bullet \end{array}\right)$} &
			\multirow{3}{*}{$L_e-L_\tau$} &
			\multirow{3}{*}{$\textrm{diag}(1,0,-1)$} &
			\multirow{3}{*}{$V_e$} &
			\multirow{3}{*}{$V^{\oplus}_{e}+V^{\leftmoon}_{e}+V^{\rm MW}_{e}+V^{\astrosun}_{e}+V^{\rm cos}_{e}$}
			\\
			& & & & &
			\\
			& & & & &
			\\
			\hline
			\multirow{6}{*}{T6} &
			\multirow{6}{*}{$\left(\begin{array}{ccc} \bullet & & \\ & 0 & \\ & & 0 \end{array}\right)$} &
			$B-3L_e$ &
			$\textrm{diag}(1,0,0)$ &
			$9V_{e}-3(V_{p}+V_{n})$ &
			$3(V^{\oplus}_{e}+V^{\leftmoon}_{e}+V^{\rm MW}_{e})+\frac{21}{4}V^{\astrosun}_{e}+\frac{39}{7}V^{\rm cos}_{e}$
			\\
			\cline{3-6}
			&
			&
			$L-3L_e$ &
			$\textrm{diag}(1,0,0)$ &
			$6V_e$ &
			$6(V^{\oplus}_{e}+V^{\leftmoon}_{e}+V^{\rm MW}_{e}+V^{\astrosun}_{e}+V^{\rm cos}_{e})$
			\\
			\cline{3-6}
			&
			&
			$B-\frac{3}{2}(L_\mu+L_\tau)$ &
			$\textrm{diag}(1,0,0)$ &
			$\frac{3}{2}(V_p+V_n)$ &
			$3(V^{\oplus}_{e}+V^{\leftmoon}_{e}+V^{\rm MW}_{e})+\frac{15}{8}V^{\astrosun}_{e}+\frac{12}{7}V^{\rm cos}_{e}$
			\\
			\cline{3-6}
			&
			&
			$L_e-\frac{1}{2}(L_\mu+L_\tau)$ &
			$\textrm{diag}(1,0,0)$ &
			$\frac{3}{2}V_e$ &
			$\frac{3}{2}(V^{\oplus}_{e}+V^{\leftmoon}_{e}+V^{\rm MW}_{e}+V^{\astrosun}_{e}+V^{\rm cos}_{e})$
			\\
			\cline{3-6}
			&
			&
			$L_e+2L_\mu+2L_\tau$ &
			$\textrm{diag}(-1,0,0)$ &
			$V_e$ &
			$V^{\oplus}_{e}+V^{\leftmoon}_{e}+V^{\rm MW}_{e}+V^{\astrosun}_{e}+V^{\rm cos}_{e}$
			\\
			\cline{3-6}
			&
			&
			$B_y+L_\mu+L_\tau$ &
			$\textrm{diag}(-1,0,0)$ &
			$V_p+V_n$ &
			$2(V^{\oplus}_{e}+V^{\leftmoon}_{e}+V^{\rm MW}_{e})+\frac{5}{4}V^{\astrosun}_{e}+\frac{8}{7}V^{\rm cos}_{e}$
			\\
			\hline
		\end{tabular}
	\end{adjustbox}
	\caption{A list of the flavor-dependent $U(1)^\prime$ gauge symmetries and corresponding matter potentials induced by them. Among the listed textures, the first three (i.e., T1, T2, and T3) are considered in this analysis while the remaining ones are included for completeness. Entries marked with $\bullet$ in matrices of the second column denote nonzero components of the $V_{\rm LRI}$ matrices. The fourth column presents the texture of $V_{\rm LRI}$, whereas the fifth column shows the form of $V_{\rm LRI}$ in terms of electron, proton, and neutron, as given by Eq.~\ref{eq:total_pot}. The last column shows the form of $V_{\rm LRI}$ in terms of the contributions from different objects in the Universe as given by Eq.~\ref{eq:total_lri}, which is later used to translate bound on $V_{\rm LRI}$ into that on $m_{Z^\prime}$ and $g^\prime$.}
	\label{tab:symm}
\end{table*}

Among these anomaly-free combinations, the flavor-universal symmetries, such as $B-L$, affect all neutrino flavors identically, and therefore, do not modify neutrino oscillation probabilities. We therefore focus on flavor-dependent $U(1)^\prime$ symmetries, which induce non-universal interactions, leading to observable modifications in neutrino flavor evolution~\cite{Davoudiasl:2011sz, Araki:2012ip, Coloma:2020gfv,delaVega:2021wpx,Agarwalla:2024ylc}. We summarize all the candidate flavor-dependent $U(1)^\prime$ symmetries in Table~\ref{tab:symm}, where we have classified them according to the texture of their induced matter potential, $\mathbf{V}_{\rm LRI}$ (see Ref.~\cite{Agarwalla:2024ylc} for a detailed discussion on $U(1)^\prime$ symmetries). This classification is particularly useful because different symmetries can lead to identical modifications in the effective Hamiltonian, giving rise to similar effects on neutrino oscillation probabilities. At the same time, distinct textures lead to characteristic signatures in neutrino oscillations.

Our analysis is based on a high-purity $\nu_\mu$ CC event sample from DeepCore, whose sensitivity is primarily driven by the $\nu_\mu \rightarrow \nu_\mu$ disappearance channel. This channel enables us to explore the symmetries, which directly impact the $\nu_\mu$ and $\nu_\tau$ sectors. Therefore, in this work, we consider only the first three textures (T1, T2, and T3) listed in the first three rows of Table~\ref{tab:symm} corresponding to the symmetries $B-3L_\mu, \,L-3L_\mu, \,B-3L_\tau, \,L-3L_\tau, \,L_\mu-L_\tau,$ and $B-L_e-2L_\tau$. The symmetry textures T4 and T5 corresponding to $L_e - L_\mu$ and $L_e - L_\tau$, respectively, have already been studied in our previous work~\cite{Garg:2026gwx} with the same dataset. As far as the remaining texture T6 is concerned, it predominantly affects only the $\nu_e$ sector (see lower panels in Fig.~\ref{fig:osc_prob}). Therefore, we do not consider texture T6 in the present study.

Gauging any of these symmetries introduces a new vector gauge boson, $Z^\prime$, mediating flavor-dependent neutral-current interactions between neutrinos and matter in addition to the standard weak interactions. We can write the effective Lagrangian for a given $U(1)^\prime$ symmetry as, 
\begin{equation}\label{eq:lagrangian}
	\mathcal{L} = \mathcal{L}_{Z^\prime}+\mathcal{L}_{\rm mix}\,, 
\end{equation}	
where the first term, $\mathcal{L}_{Z^\prime}$, describes the new interaction mediated by $Z^\prime$ with coupling strength $g_{Z^\prime}$~\cite{He:1990pn,He:1991qd,Heeck:2010pg} as,
\begin{equation}
	\begin{aligned}
		\mathcal{L}_{Z^\prime} = -g_{Z^\prime} \Big(
		& a_e \, \bar{e} \gamma^\alpha e
		+ a_u \, \bar{u} \gamma^\alpha u
		+ a_d \, \bar{d} \gamma^\alpha d
		+ b_e \, \bar{\nu}_e \gamma^\alpha P_L \nu_e \\
		& + b_\mu \, \bar{\nu}_\mu \gamma^\alpha P_L \nu_\mu
		+ b_\tau \, \bar{\nu}_\tau \gamma^\alpha P_L \nu_\tau
		\Big) Z^\prime_\alpha \,.
	\end{aligned}
\end{equation}
Here $a_u$, $a_d$, and $a_e$ correspond to $U(1)^\prime$ charges of the up quark, down quark, and electron, respectively, while $b_e$, $b_\mu$, and $b_\tau$ correspond to that for the three neutrino flavors (See Appendix~\ref{app:U1-charges} for the charge assignments). The second term in Eq.~\ref{eq:lagrangian} accounts for the mixing between $Z$ and $Z^\prime$~\cite{Heeck:2010pg,Joshipura:2019qxz,Babu:1997st}, which is parameterized by $(\xi - \sin\theta_W\chi)$, where $\chi$ denotes the kinetic mixing angle, and $\xi$ represents the rotation angle between the gauge and mass eigenstates. This mixing induces an effective interaction between neutrinos and matter as,
\begin{equation}
	\mathcal{L}_{\rm mix}
	= -g_{Z^\prime} \,\frac{e}{\sin\theta_W \cos\theta_W}\,
	(\xi - \sin\theta_W \chi)\,
	J^{\sigma \prime}\, J^{3}_{\sigma} \,,
\end{equation}
where,
\begin{align}
	J^{\sigma \prime} &= \bar{\nu}_\mu \gamma^\sigma P_L \nu_\mu 
	- \bar{\nu}_\tau \gamma^\sigma P_L \nu_\tau \, , \\
	J^{3}_{\sigma} &= -\tfrac{1}{2}\,\bar{e}\gamma_\sigma P_L e 
	+ \tfrac{1}{2}\,\bar{u}\gamma_\sigma P_L u 
	- \tfrac{1}{2}\,\bar{d}\gamma_\sigma P_L d \,.
\end{align}
Here, $e$ is the electric charge, $\theta_W$ is the Weinberg angle, and $P_L$ is the left-handed projection operator. In this $Z-Z^\prime$ mixing scenario, electron and proton contributions cancel each other, leaving neutrons as the dominant source of the induced potential. Since the value of the mixing parameter $(\xi - \sin\theta_W \chi)$ is not precisely known, we express our results in terms of the effective coupling $g_{Z^\prime}(\xi - \sin\theta_W \chi)$ for the $Z-Z^\prime$ mixing scenario.

For purely leptonic symmetries, the matter potential is predominantly sourced by electrons due to the absence of muons and taus in ordinary matter. In contrast, the symmetries involving both baryon and lepton numbers receive contributions from electrons, protons, and neutrons. The LRI potential experienced by a neutrino at a distance $d$ from a source containing $N_f$ number of fermions $f$ (\textit{i.e.}, $e, p, n$) due to $Z^\prime$ exchange follows the Yukawa-like potential as,
\begin{equation}\label{eq:V_Z}
	V_{f} = N_f\frac{g^{\prime 2}}{4\pi d} \, e^{-m_{Z^\prime} d},
\end{equation}
where $m_{Z^\prime}$ is the mediator mass. For the $L_\mu - L_\tau$ symmetry, the direct Yukawa contribution in Eq. \ref{eq:V_Z} is absent due to the lack of muons and taus in matter. Instead, the dominant contribution arises from $Z$--$Z^\prime$ mixing sourced by $N_n$ number of neutrons as,
\begin{equation}\label{eq:V_ZZ'}
	V_{f} = N_n\frac{g^{\prime 2}}{4\pi d}\,\frac{e}{\sin\theta_W \cos\theta_W \,}  e^{-m_{Z^\prime} d} \,.
\end{equation}
In Eqs.~\ref{eq:V_Z} and \ref{eq:V_ZZ'}, we define the effective coupling as,
\begin{equation}
	g^\prime =
	\begin{cases}
		g_{Z^\prime} & \text{(via $Z^\prime$ only )}\,,\\[6pt]
		\sqrt{g_{Z^\prime}\,(\xi - \sin\theta_W \chi)} & \text{(via $Z$--$Z^\prime$ mixing)}\,.
	\end{cases}
\end{equation}

The total LRI potential sourced from a given fermion is obtained by summing over the contributions from all relevant astrophysical sources, including the Earth ($\oplus$), Moon($\leftmoon$), Sun ($\odot$), Milky Way (MW), and cosmological matter distribution (cos) following Refs.~\cite{Bustamante:2018mzu,Singh:2023nek,Agarwalla:2024ylc} as,
\begin{equation}
	V_f(m_{Z^\prime}, g^\prime) = V_f^{\oplus} + V_f^{\leftmoon} + V_f^{\odot} + V_f^{\mathrm{MW}} + V_f^{\mathrm{cos}} \, .
\end{equation}
Therefore, the resulting LRI potential for a neutrino of flavor $\alpha \in \{e,\mu,\tau\}$ is,
\begin{equation}\label{eq:total_pot}
	V_{\mathrm{LRI},\alpha}(m_{Z^\prime}, g^\prime) = b_\alpha \sum_{f=e,p,n} \kappa_f \, V_f(m_{Z^\prime}, g^\prime) \, ,
\end{equation}
where $b_\alpha$ is the $U(1)^\prime$ charge of $\nu_\alpha$. For symmetries other than $L_\mu - L_\tau$, $\kappa_f \equiv a_f$ correspond to the $U(1)^\prime$ charges of the fermions, \textit{i.e.}, $a_e$ for electrons, $a_p = 2a_u + a_d$ for protons, and $a_n = 2a_d + a_u$ for neutrons. In case of $L_\mu - L_\tau$, the interaction arises from $Z$--$Z^\prime$ mixing and $\kappa_f$ is replaced by the weak hypercharge $y_n = 2y_d + y_u$ for neutrons.

The relative contributions from different sources depend on the interaction range governed by $1/m_{Z^\prime}$, beyond which the potential is exponentially suppressed.	We follow Ref.~\cite{Agarwalla:2024ylc} to calculate the contributions from different astronomical sources. As far as the number densities of electron ($N_e$), proton ($N_p$) and neutron ($N_n$) are concerned, we assume electrically neutral ($N_e = N_p$), and approximately isoscalar ($N_p = N_n$) matter distributions with appropriate corrections for different sources. We consider the Sun~\cite{Heeck:2010pg} ($N_{e,\odot} = N_{p,\odot} \sim 10^{57}$ and $N_{n,\odot} \approx N_{e,\odot}/4$) and Moon ($N_{e,\leftmoon} = N_{p,\leftmoon} = N_{n,\leftmoon} \sim 5\times10^{49}$) as point-like sources. In contrast, the Earth ($N_{e,\oplus} = N_{p,\oplus} = N_{n,\oplus} \sim 4\times10^{51}$), Milky Way ($N_{e,\mathrm{MW}} = N_{p,\mathrm{MW}} \approx N_{n,\mathrm{MW}} \sim 10^{67}$), and cosmological matter distribution ($N_{e,\mathrm{cos}} = N_{p,\mathrm{cos}} \sim 10^{79}$ and $N_{n,\mathrm{cos}} \sim 10^{78}$) are modeled as continuous matter distributions.

We evaluate an average LRI potential at the detector location, neglecting the spatial variations of the potential along the neutrino trajectory inside the Earth following Refs.~\cite{Bustamante:2018mzu,Agarwalla:2023sng,Singh:2023nek,Agarwalla:2024ylc}. This approximation is valid when the interaction range exceeds the radius of the Earth ($m_{Z^\prime} \lesssim 10^{-14}\,\mathrm{eV}$) ensuring coherent contributions from all matter constituents~\cite{Wise:2018rnb}. With these assumptions, the total LRI potential can be computed by evaluating only the electron contribution explicitly and incorporating proton and neutron effects through rescaling in terms of the potential from electrons as,
\begin{equation}\label{eq:total_lri}
	\begin{aligned}
	V_{\mathrm{LRI},\alpha} = & b_\alpha \left[ \kappa_e 
	+ \kappa_p \frac{N_{p,\oplus}}{N_{e,\oplus}} 
	+ \kappa_n \frac{N_{n,\oplus}}{N_{e,\oplus}} \right] V_{e}^{\oplus}
	+ (\oplus \rightarrow \leftmoon) \\
	& + (\oplus \rightarrow \odot) + (\oplus \rightarrow \mathrm{MW}) + (\oplus \rightarrow \mathrm{cos}) \, ,
	\end{aligned}
\end{equation}
where the notation $(\oplus \rightarrow X)$ denotes analogous contributions from the Moon, Sun, Milky Way, and the cosmological matter distribution. The resulting total potentials for each symmetry using Eq.~\ref{eq:total_lri} are summarized in Table~\ref{tab:symm} and are used to translate bounds from $V_{\mathrm{LRI}}$ to constraints on the coupling and mass of the mediator as shown in Fig. ~\ref{fig:g_m_bound}.

\section{Effect of LRI on neutrino oscillations}
\label{sec:Hamiltonian}

The presence of long-range interactions, in addition to the standard interactions (SI), modifies neutrino propagation through an additional flavor-dependent matter potential. The effective Hamiltonian in the flavor basis can be written as,
\begin{equation}
	H_f=U\left[
	\begin{tabular}{c c c} 0 & 0 & 0\\
		0 & $\frac{\Delta m^2_{21}}{2E}$ & 0\\
		0 & 0 & $\frac{\Delta m^2_{31}}{2E}$
	\end{tabular}\right]U^\dag\, + \left[
	\begin{tabular}{c c c} $V_{\rm CC}$ & 0 & 0\\ 0 & 0 & 0 \\ 0 & 0 & 0
	\end{tabular} \right] + \left[
	\begin{tabular}{c c c} $\zeta$ & 0 & 0 \\ 0 & $\xi$ & 0\\ 
		0 & 0 & $\eta$ \\
	\end{tabular} \right],\
	\label{eq:modified_H}
\end{equation}
where the first term corresponds to vacuum oscillations governed by the Pontecorvo-Maki-Nakagawa-Sakata (PMNS) mixing matrix $U$ and the mass-squared differences $\Delta m_{ij}^2$. The second term represents the standard matter effect with $V_{\rm CC} = \pm \sqrt{2}\,G_F N_e$, arising from charged-current interactions of neutrinos (positive) and antineutrinos (negative) with electrons in the Earth, where $G_F$ is the Fermi constant and $N_e$ is the electron number density.

The third term diag($\zeta, \xi, \eta$) characterize LRI contribution to the Hamiltonian for the underlying $U(1)^\prime$ symmetry textures given in Table~\ref{tab:symm}, where the corresponding LRI potential is obtained from Eq.~\ref{eq:total_lri}. In this work, we consider three representative textures, T1 : diag$(0, -1, 0)$, T2 : diag$(0, 0, -1)$, and T3 : diag$(0, 1, -1)$, as listed in Table~\ref{tab:symm}.

\subsection{Effect of LRI on neutrino oscillation probabilities}
\label{sec:prob_var_pot}
The neutrino transition probability $P({\nu_\alpha \to \nu_\beta})$ in the presence of both SI and LRI is governed by the effective Hamiltonian in Eq.~(\ref{eq:modified_H}). In this work, we compute neutrino oscillation probabilities assuming normal mass ordering (NO) using a 12-layered density profile based on the Preliminary Reference Earth Model (PREM)~\cite{Dziewonski:1981xy}. The calculations are performed using the publicly available \texttt{PISA} framework developed by the IceCube Collaboration~\cite{IceCube:2018ikn}. The presence of LRI with different textures modify the effective mixing parameters and mass splittings in different ways, leading to characteristic distortions in the oscillation patterns. In general, the precise features depend on the flavor structure of the LRI potential, with sizable deviations arising when the LRI contribution becomes comparable to $\Delta m^2_{31}/2E$ and $V_{\rm CC}$ with values of the order of $\sim 10^{-13}$ eV.

\begin{figure}
	\centering
	\includegraphics[width=\linewidth]{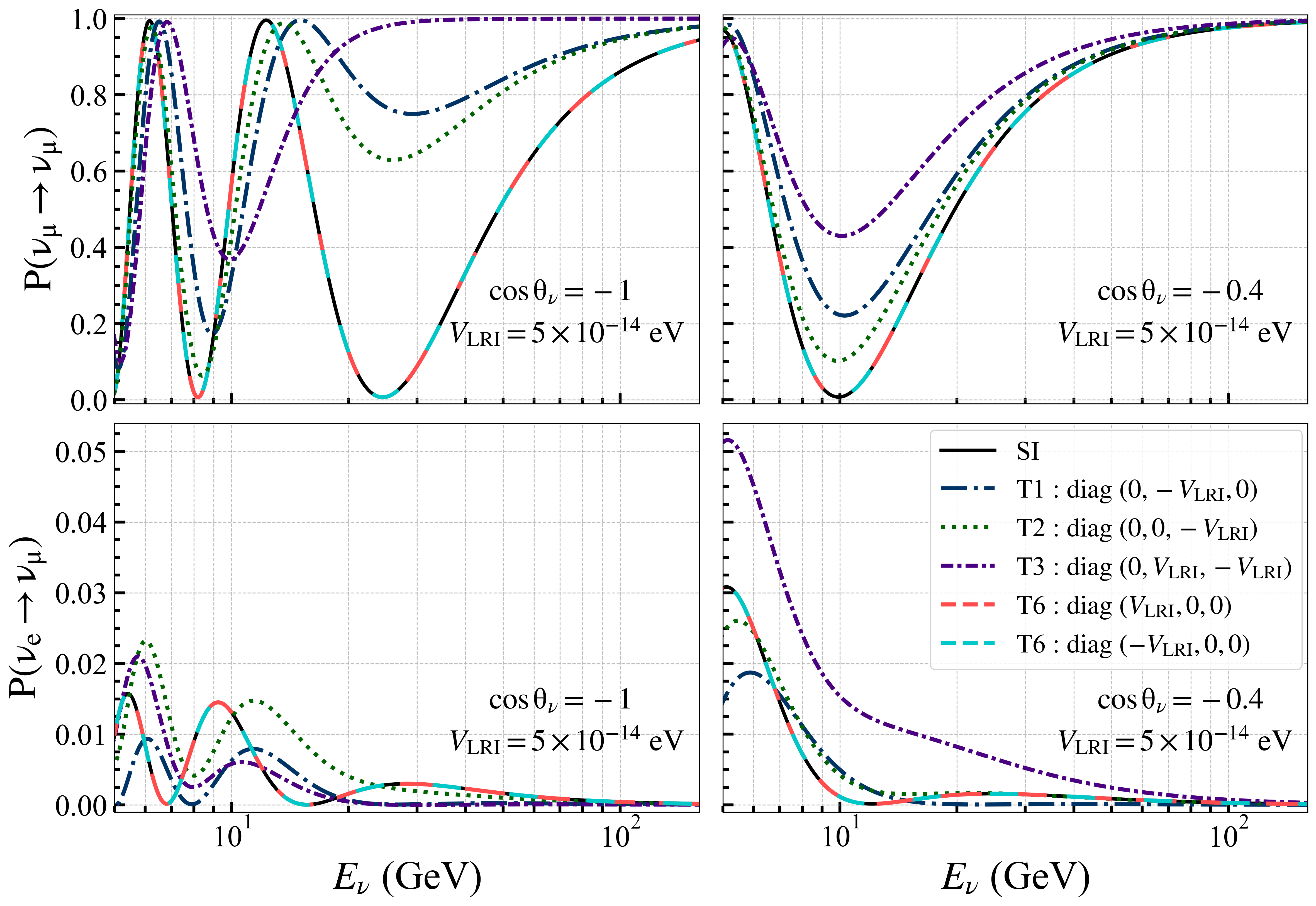}
	\caption{The oscillation probabilities $P(\nu_\mu \rightarrow \nu_\mu)$ and $P(\nu_e \rightarrow \nu_\mu)$ are shown as functions of neutrino energy in the top and bottom panels, respectively, for SI and SI + LRI with different textures listed in Table~\ref{tab:symm}. The left and right panels correspond to $\cos\theta_\nu$ of $-1$ and $-0.4$, respectively. For the LRI scenarios, we choose a representative value of $V_{\mathrm{LRI}} = 5 \times 10^{-14}~\mathrm{eV}$. We assume NO with $\theta_{23} = 45.57^\circ$ and $\Delta m^2_{31} = 2.48 \times 10^{-3}~\mathrm{eV}^2$.}
	\label{fig:osc_prob} 
\end{figure}

The $\nu_\mu$ CC events at DeepCore mainly come from $\nu_\mu \rightarrow \nu_\mu$ survival channel with a small contribution from $\nu_e \rightarrow \nu_\mu$ appearance channel. Therefore, to illustrate the impact of different LRI textures listed in Table~\ref{tab:symm} on neutrino oscillations, we present $\nu_\mu \rightarrow \nu_\mu$ disappearance probability in the top row and $\nu_e \rightarrow \nu_\mu$ appearance probability in the bottom row of Fig.~\ref{fig:osc_prob}. The left and right panels correspond to neutrinos traversing the Earth along trajectories with $\cos\theta_\nu = -1$ (core-passing) and $\cos\theta_\nu = -0.4$ (only mantle-passing), respectively. In each panel, we compare the effects of the representative LRI textures of T1, T2, T3, and T6. For the value of LRI potential, we consider a representative choice of $V_{\rm LRI} = 5 \times 10^{-14}\,\text{eV}$. In Fig.~\ref{fig:osc_prob}, we can observe that the impact of LRI is more pronounced for neutrinos traversing longer trajectories ($\cos\theta_\nu = -1$) as compared to the shorter trajectories ($\cos\theta_\nu = -0.4$).

Among all the textures shown in Fig.~\ref{fig:osc_prob}, the texture T3 produces the largest deviations in the $\nu_\mu \rightarrow \nu_\mu$ channel, as it directly affects the $\nu_\mu$ and $\nu_\tau$ sectors, which is expected to give rise to stronger constraints on the corresponding LRI potential. The textures T2 and T3 also show a sizable amount of difference from the standard interaction curve. These modifications at the oscillation probability level derives observable distortions in the reconstructed event distributions at DeepCore, which forms the basis of LRI sensitivities in this study. As far as, both the possible configurations of T6 are concerned, they exhibit negligible impact on the $\nu_\mu \rightarrow \nu_\mu$ channel, as they primarily modify the electron sector, and therefore, affect appearance channel $\nu_e \rightarrow \nu_\mu$. This explains the limited sensitivity of a $\nu_\mu$ dominated dataset to such scenarios and motivates their exclusion from further analysis.

\subsection{Effect of LRI on neutrino oscillograms}
\label{sec:prob_oscillogram_var_pot}
\begin{figure*}
	\centering
	\includegraphics[width=\textwidth]{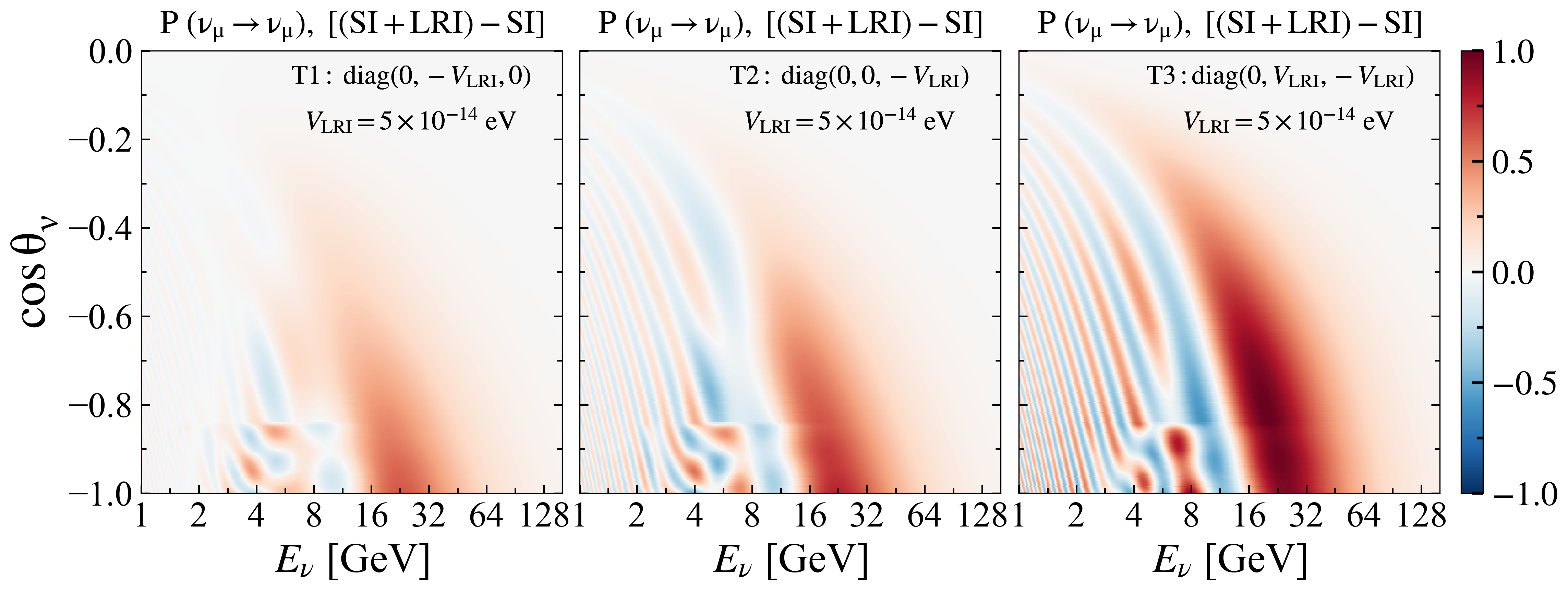}
	\caption{The difference between the $P(\nu_\mu \rightarrow \nu_\mu)$ oscillograms in the $(E_\nu,\, \cos\theta_\nu)$ plane for the SI + LRI ($V_{\rm LRI} = 5 \times 10^{-14}~\mathrm{eV}$) and SI scenarios. The left, middle, and right panels correspond to textures, T1, T2, and T3, respectively. We assume NO with $\theta_{23} = 45.57^\circ$ and $\Delta m^2_{31} = 2.48 \times 10^{-3}~\mathrm{eV}^2$.}
	\label{fig:oscillogram}
\end{figure*}

Figure~\ref{fig:oscillogram} illustrates the impact of LRI on the oscillation probability for the $\nu_\mu \rightarrow \nu_\mu$ disappearance channel in the $(E_\nu,\,\cos\theta_\nu)$ plane. The figure shows the difference between probabilities for the SI+LRI and SI scenarios for three representative textures T1, T2, and T3, displayed in the left, middle, and right panels, respectively. The value of LRI potential is taken as a representative choice of $V_{\rm LRI} = 5 \times 10^{-14}\,\text{eV}$.

In Fig.~\ref{fig:oscillogram}, the effect of LRI is clearly visible as a modification of the characteristic oscillation valley (red band), particularly at longer baselines ($\cos\theta_\nu < -0.7$) and in the mid-energy range of around 16 GeV to 32 GeV across all panels. Notably, the T3 texture exhibits the most pronounced deviation, as seen by the enhanced red region in the right panel. This behavior is consistent with the trends observed for the oscillation probability for a given baseline in Fig.~\ref{fig:osc_prob}. For completeness, we provide the probability oscillograms for the SI and SI + LRI scenarios for all three textures in Appendix~\ref{app:Oscillograms}. We now proceed to describe the IceCube DeepCore detector.

\section{IceCube DeepCore}
\label{sec:DeepCore}
IceCube is a cubic-kilometer neutrino observatory situated at the geographic South Pole~\cite{IceCube:2016zyt}. The detector is composed of 5,160 digital optical modules (DOMs)~\cite{IceCube:2010dpc}, installed on 86 vertical strings that are embedded deep within the Antarctic ice at depths ranging from 1,450 m to 2,450 m. Each DOM houses a photomultiplier tube (PMT) and readout electronics capable of detecting the faint Cherenkov light produced by relativistic charged particles traversing the ice. These charged particles are generated during neutrino interactions with the ice, allowing IceCube to infer the energy, arrival direction, and flavor of the incoming neutrinos. IceCube is particularly sensitive to TeV - PeV energy astrophysical neutrinos.

The bottom-central region of IceCube, known as DeepCore~\cite{IceCube:2011ucd}, is a densely instrumented sub-array designed for the detection of GeV-energy atmospheric neutrinos. DeepCore consists of seven additional strings outfitted with closely-spaced high-quantum-efficiency DOMs, which lowers the energy threshold to a few GeV - ideal for studying atmospheric neutrino oscillations. Exploiting these features, DeepCore has provided the precision measurements of atmospheric mixing angle $\theta_{23}$ and the mass-squared difference $\Delta m^2_{32}$~\cite{IceCube:2014flw,IceCube:2017lak,IceCubeCollaboration:2023wtb,IceCubeCollaboration:2024ssx}, and observed the tau neutrino appearance~\cite{IceCube:2019dqi}. Furthermore, DeepCore can also perform searches for several BSM physcis scenarios, including NSI, sterile neutrinos, flavor-dependent LRI, decoherence, and quantum gravity etc~\cite{IceCube:2017ivd,IceCube:2020phf,IceCube:2020tka, IceCubeCollaboration:2024nle,IceCube:2024pky,IceCube:2024dlz, IceCube:2017zcu,IceCubeCollaboration:2021euf,IceCube:2025kve}. In the present analysis, we use GeV-energy atmospheric neutrinos observed at DeepCore to search for long-range neutrino interactions in the context of flavor-dependent $U(1)^\prime$ symmetries.

\section{Events at IceCube DeepCore}
\label{sec:events}

In this work, we utilize the publicly available atmospheric neutrino events collected by IceCube DeepCore between 2011 and 2019, corresponding to a total livetime of 7.5 years~\cite{IceCubeCollaboration:2023wtb,DVN/B4RITM_2025}. The sample contains 21,914 neutrino events within the reconstructed energy interval of 6.3 GeV to 158.5 GeV. This dataset, commonly referred to as the “golden event sample,” is constructed by rejecting DOM hits associated with highly scattered photons, thereby improving the quality of event reconstruction. Compared to earlier data releases, this sample incorporates several important improvements, including refined calibration of the optical modules using in-situ data~\cite{IceCube:2020nwx}, a more accurate description of the Antarctic ice properties~\cite{Chirkin:2013tma}, updated Monte Carlo (MC) simulations, and more efficient event selection techniques with enhanced background rejections. Additionally, improved reconstruction techniques~\cite{IceCube:2022kff} and a more sophisticated handling of systematic uncertainties~\cite{IceCubeCollaboration:2023wtb} contribute to the overall reliability of the dataset.

This analysis is carried out using a forward-folding strategy based on detailed MC simulations provided by the IceCube Collaboration~\cite{DVN/B4RITM_2025}. Simulated events are reweighted within the PISA framework~\cite{IceCube:2018ikn} to incorporate the effects of atmospheric neutrino fluxes, interaction cross sections, neutrino oscillations, and detector response. This procedure ensures that the expected MC event distribution closely matches the observed event distribution at IceCube DeepCore. Identical filtering and reconstruction algorithms developed using simulations are applied to both expected MC and observed data.

To reduce background contamination from atmospheric muons and detector noise, a series of selection cuts is employed, yielding a dataset dominated by neutrino-induced events. The reconstructions of key observables, such as the neutrino energy ($E_{\rm reco}$) and cosine of the zenith angle ($\cos\theta_{\rm reco}$), are performed using likelihood-based methods~\cite{IceCube:2014flw,IceCube:2022kff,Garza2014Measurement,AndriiThesis}. The events are classified into topologies based on their spatial and temporal patterns of light deposition in the detector, which are indicative of the underlying neutrino flavor and interaction type. This classification is achieved through a Boosted Decision Tree (BDT)~\cite{Friedman:2001wbq}, which assigns a particle identification (PID) score indicating the probability that an event is from a $\nu_{\mu}$ CC interaction. The $\nu_{\mu}$ CC interaction produces a muon, which travels a long distance in the ice and deposits energy along its path resulting in a long elongated signal pattern, thus are referred to as track-like events. On the other hand, $\nu_e$ CC, $\nu_\tau$ CC, and neutral-current (NC) interactions of all flavors, deposit their energy in a very short distance, which results in a spherical pattern of light, thus are referred to as cascade-like events. The PID score is a continuous value between 0 and 1, where higher values indicate high probability for the event being track-like. For this analysis, neutrinos and antineutrinos are treated together because their event topologies are nearly indistinguishable in DeepCore.

In this work, the events are binned in 10 logarithmic bins in $E_{\rm reco}$ from 6.3 GeV to 158.5 GeV, and 10 linearly-spaced bins in $\cos\theta_{\rm reco}$ from $-1$ to $0.1$. The highest-energy bin is widened to maintain adequate statistics. Events are further divided into two categories based on their PID: mixed events for PID $\in [0.55,\, 0.75]$ and track-like events for PID $\in [0.75, 1.0]$. Events with $\text{PID} < 0.55$, which are predominantly cascade-like, are excluded by the IceCube Collaboration to obtain this high-purity $\nu_\mu$ CC event sample.

\begin{figure}[t]
	\centering
	\includegraphics[width=\linewidth]{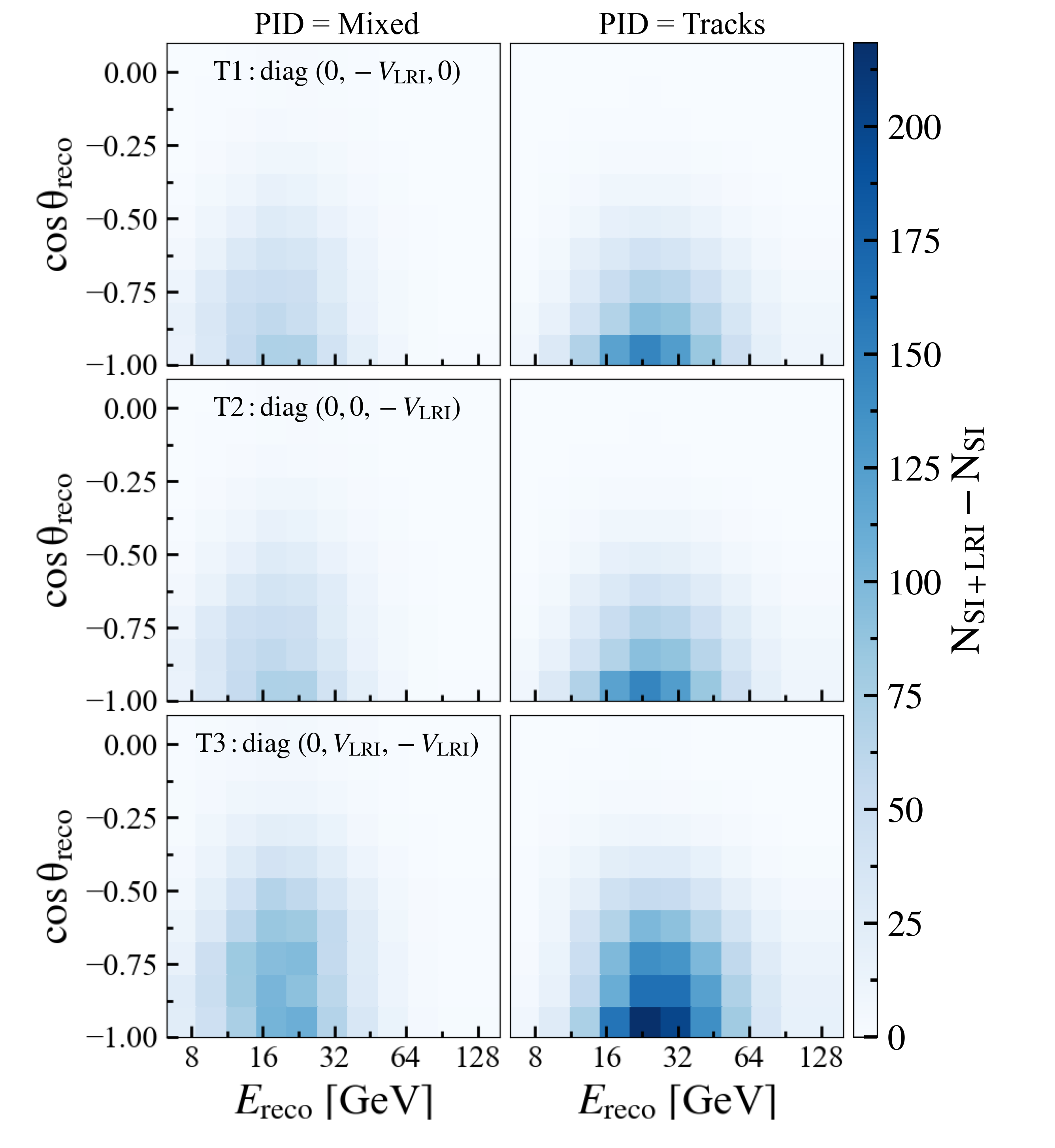}
	\caption{The difference between the expected MC events at DeepCore for the SI + LRI ($V_{\rm LRI} = 5 \times 10^{-14}~\mathrm{eV}$) and SI scenarios. The top, middle, and bottom panels correspond to the three textures, T1, T2, and T3, respectively. The left and right panels show the mixed and track-like events, respectively. We use the nominal values of the nuisance parameters as listed in Table~\ref{tab:systematic_params} in the Appendix and assume NO with $\theta_{23} = 45.57^\circ$ and $\Delta m^2_{31} = 2.48 \times 10^{-3}~\mathrm{eV}^2$.}
	\label{fig:event_diff} 
\end{figure}

To illustrate the LRI effect at the reconstructed event level, we compare the predicted event distributions for the SI case with those for the SI+LRI case. Figure~\ref{fig:event_diff} shows the event differences between SI+LRI and SI as a function of $E_{\rm reco}$ and $\cos\theta_{\rm reco}$ for T1, T2, and T3 textures in the top, middle, and bottom panels, respectively. The mixed and track-like events are shown in the left and right panels, respectively. The largest deviations in the event counts occur in the regions corresponding to the longer baselines and intermediate neutrino energies. Among the three textures, T3 produces the most pronounced effects in the event differences. These features are consistent with the behavior observed at the oscillation probability level in Figs.~\ref{fig:osc_prob} and \ref{fig:oscillogram}.

\section{Limits on LRI potential}
\label{sec:results}

We derive the constraints on the LRI potential ($V_{\rm LRI}$) associated with the $U(1)^\prime$ symmetries by comparing the expected event distributions with 8 years of high-purity $\nu_\mu$ CC atmospheric data from IceCube DeepCore. Now we describe the statistical methods used in this analysis.

\subsection{Numerical methods}
\label{sec:stat_methods}
We perform a binned likelihood analysis using the frequentist approach, where we compare the predicted event distributions with the observed data binned in $E_{\rm reco}$, $\cos\theta_{\rm reco}$, and PID. In this work, modified $\chi^2$ is used as a test statistic following Ref.~\cite{IceCubeCollaboration:2023wtb},
\begin{equation}
	\chi^2_{\rm mod} =
	\sum_{i \in \text{bins}}
	\frac{\left(N_i^{\rm exp} - N_i^{\rm obs}\right)^2}
	{N_i^{\rm exp} + \left(\sigma_i^{\rm sim}\right)^2}
	+
	\sum_j \frac{(s_j - \hat{s}_j)^2}{\sigma_{s_j}^2}\,,
\end{equation}
where $N_i^{\rm exp}$ and $N_i^{\rm obs}$ represent the predicted and observed number of events in the  $i^{\rm th}$ analysis bin, respectively. The term $\sigma_i^{\rm sim}$ accounts for the statistical uncertainty associated with the limited size of the MC sample used to estimate the expected events. The second term acts as a prior penalty for the systematic parameters, for which the external constraints are available. For each of these systematic parameters, we consider a Gaussian prior of $\sigma_{s_j}$ centered around the nominal value of $\hat{s}_j$.

\subsection{Systematic uncertainties}
\label{sec:systs}
In the fit, we include systematic uncertainties related to the atmospheric neutrino flux, neutrino-nucleon interaction cross sections, detector response, and oscillation parameters, along with separate normalization uncertainties for neutrinos and atmospheric muon background. The treatment of systematic uncertainties in this analysis follows the detailed methodology presented in Ref.~\cite{IceCubeCollaboration:2023wtb}. Table~\ref{tab:systematic_params} in the appendix provides a complete list of the nuisance parameters with their nominal values, $1\sigma$ priors (if available), and allowed ranges for the fitting.

The detector-related uncertainties are incorporated in the form of five nuisance parameters. The overall photon detection efficiency of the DOMs is modeled via a global scaling factor. In addition, the optical properties of the refrozen ice surrounding the DOMs, distinct from the bulk glacial ice, are parameterized in terms of the two parameters, $p_0$ and $p_1$, which describe the modifications of the angular acceptance of the modules~\cite{IceCube:2023ahv}. The bulk ice itself is characterized by the two additional parameters that scale the scattering and absorption lengths, which capture the uncertainties in light propagation through the detector medium.

For the atmospheric neutrino flux, we adopt the prediction by Honda \textit{et al.}~\cite{Honda:2015fha} as the reference model. The uncertainties arising from hadron production are incorporated following the prescription in Ref.~\cite{Barr:2006it}, with calculations performed using MCEq~\cite{Fedynitch:2018cbl}. These uncertainties are parameterized in terms of the effective variations in pion and kaon productions, along with a tilt parameter that accounts for the uncertainties in the spectral index of the flux. Additionally, independent normalization factors are introduced for neutrino-induced events and atmospheric muon backgrounds.

Neutrino interaction uncertainties are treated using the GENIE event generator as the baseline model for neutrino-nucleon cross sections. The axial mass parameters, $M_A^{\rm CCQE}$ and $M_A^{\rm RES}$, which govern charged-current quasi-elastic and resonant interactions, respectively, are included as nuisance parameters, and allowed to vary within their respective $2\sigma$ ranges. For the deep-inelastic scattering (DIS), we employ a model that interpolates between the GENIE prediction~\cite{GENIE-cross} and the CSMS calculation~\cite{Cooper-Sarkar:2011jtt}. This interpolation is controlled by a parameter denoted as DIS CSMS, where a value of 0 corresponds to the GENIE model and a value of 1 corresponds to the CSMS prediction. The NC-interaction uncertainties are accounted by a normalization parameter defined as the ratio of NC to CC interaction cross sections.

As far as the neutrino oscillation parameters are concerned, we fix $\theta_{12} = 33.41^\circ$, $\theta_{13} = 8.58^\circ$, and $\Delta m^2_{21} = 7.41 \times 10^{-5}\,\text{eV}^2$, which correspond to their best-fit values from the global fit of NuFit~v5.2~\cite{Esteban:2020cvm}. The CP-violating phase $\delta_{\rm CP}$ is set to zero, since the present $\nu_\mu$ dataset has negligible sensitivity to this parameter. The remaining oscillation parameters, $\theta_{23}$ and $\Delta m^2_{31}$, are treated as free and are minimized over the ranges $[38^\circ,\,52^\circ]$ and $[2.0,\,3.0]\times10^{-3}\,\text{eV}^2$, respectively, assuming normal mass ordering.

\subsection{Constraints on the LRI potentials}
\label{sec:constraints_pot}

\begin{figure}
	\centering
	\includegraphics[width=\linewidth]{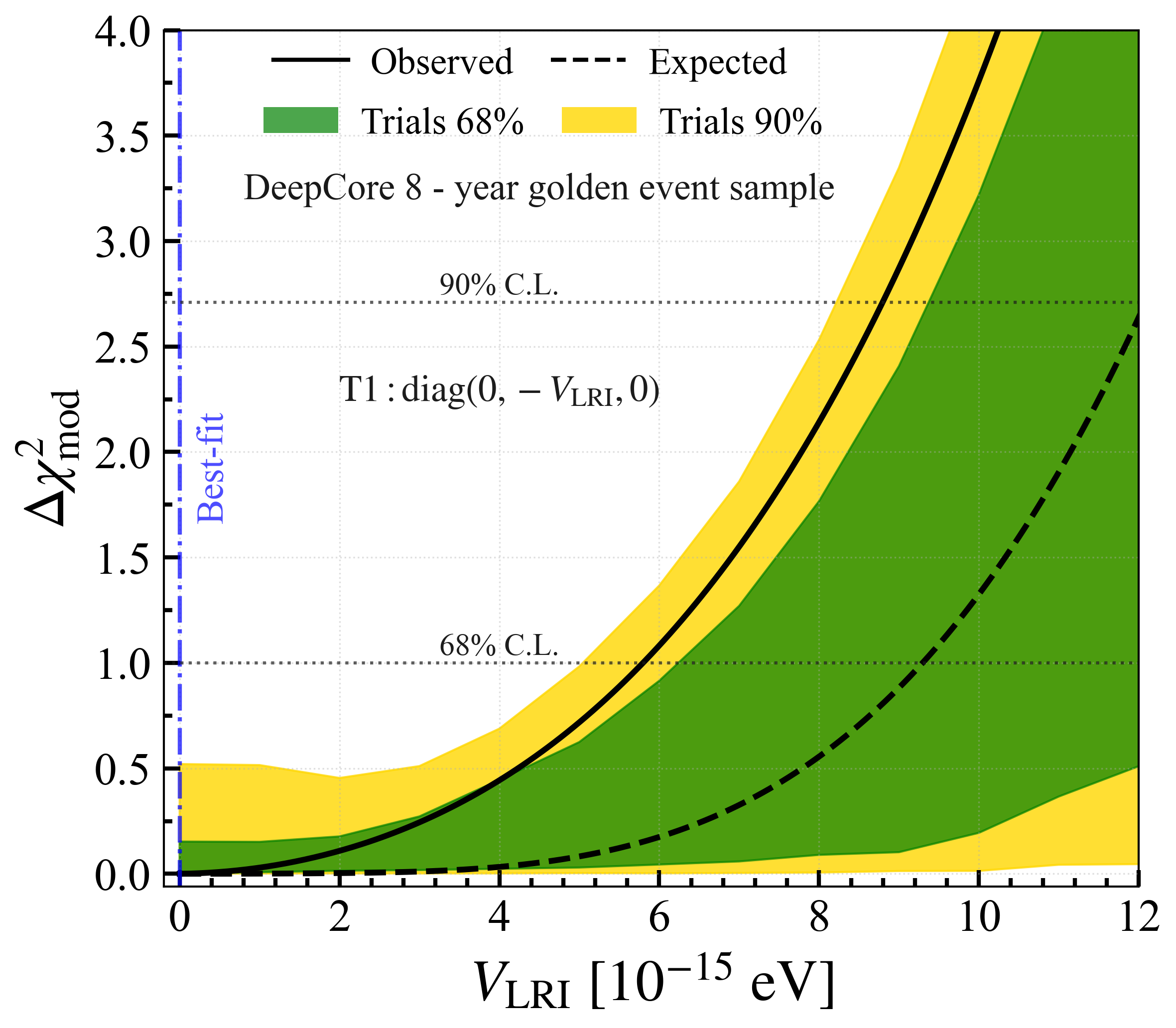}
	\caption{Constraints on the LRI potential sourced by the texture T1 of $U(1)^\prime$ symmetries using the 8-year high-purity $\nu_\mu$ CC dataset of IceCube DeepCore, assuming NO. The solid-black curve represents the observed $\Delta \chi^2_{\rm mod}$ as a function of $V_{\rm LRI}$, while the dashed-black curve shows the corresponding expected sensitivity obtained from MC. The vertical dashed-dotted blue line denote the best-fit value of $V_{\rm LRI}$. The green and yellow bands indicate the respective range of 68\% and 90\% from the distribution of $\Delta \chi^2_{\rm mod}$ which are obtained from fitting the 500 statistically fluctuated pseudo-experiments.}
		\label{fig:txt1}
\end{figure}

\begin{figure}
	\centering
	\includegraphics[width=\linewidth]{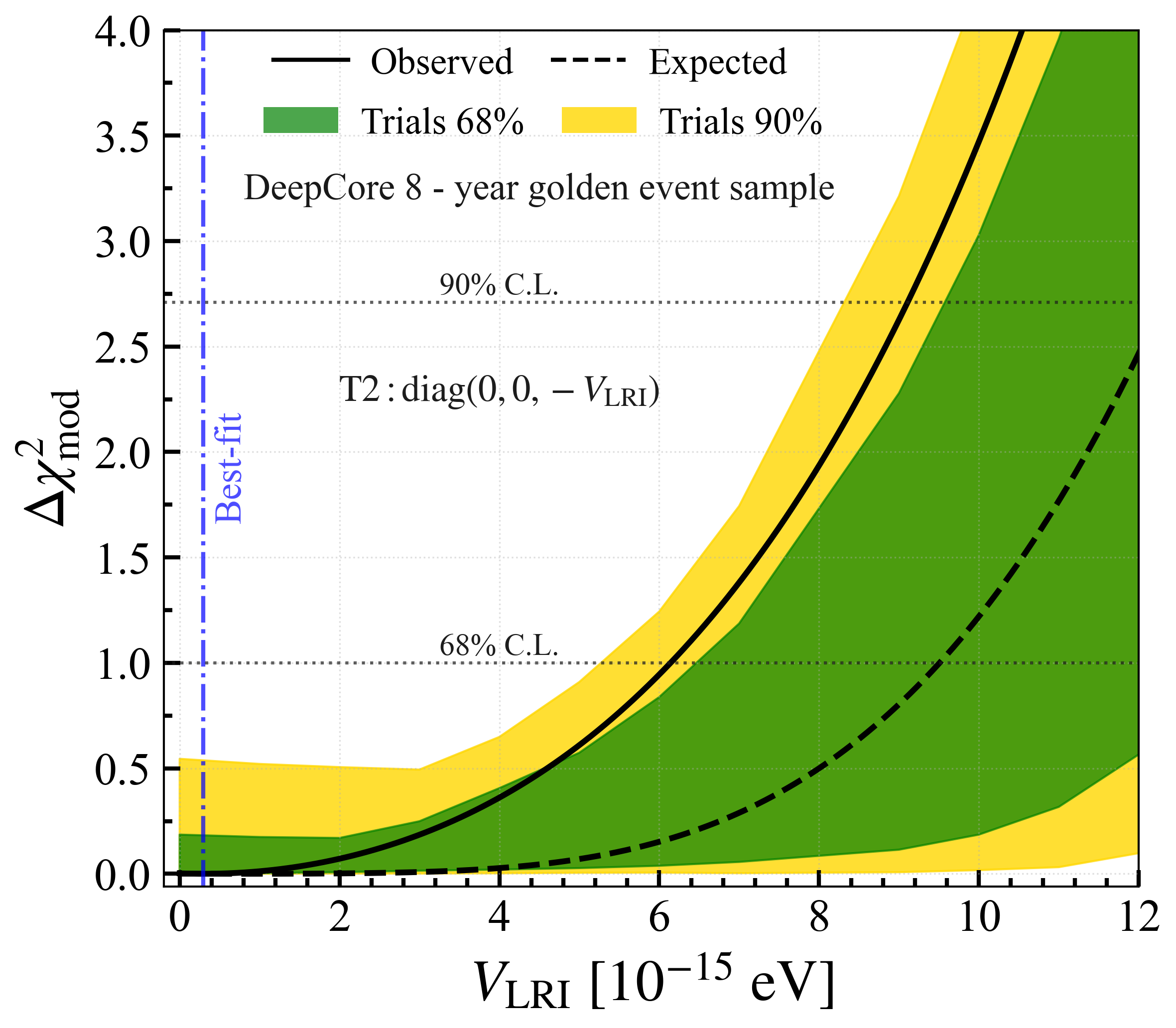}
	\caption{Constraints on the LRI potential sourced by the texture T2 of $U(1)^\prime$ symmetries using the 8-year high-purity $\nu_\mu$ CC dataset of IceCube DeepCore, assuming NO. The solid-black curve represents the observed $\Delta \chi^2_{\rm mod}$ as a function of $V_{\rm LRI}$, while the dashed-black curve shows the corresponding expected sensitivity obtained from MC. The vertical dashed-dotted blue line denote the best-fit value of $V_{\rm LRI}$. The green and yellow bands indicate the respective range of 68\% and 90\% from the distribution of $\Delta \chi^2_{\rm mod}$ which are obtained from fitting the 500 statistically fluctuated pseudo-experiments.}
		\label{fig:txt2}
\end{figure}

\begin{figure}
 \centering
 \includegraphics[width=\linewidth]{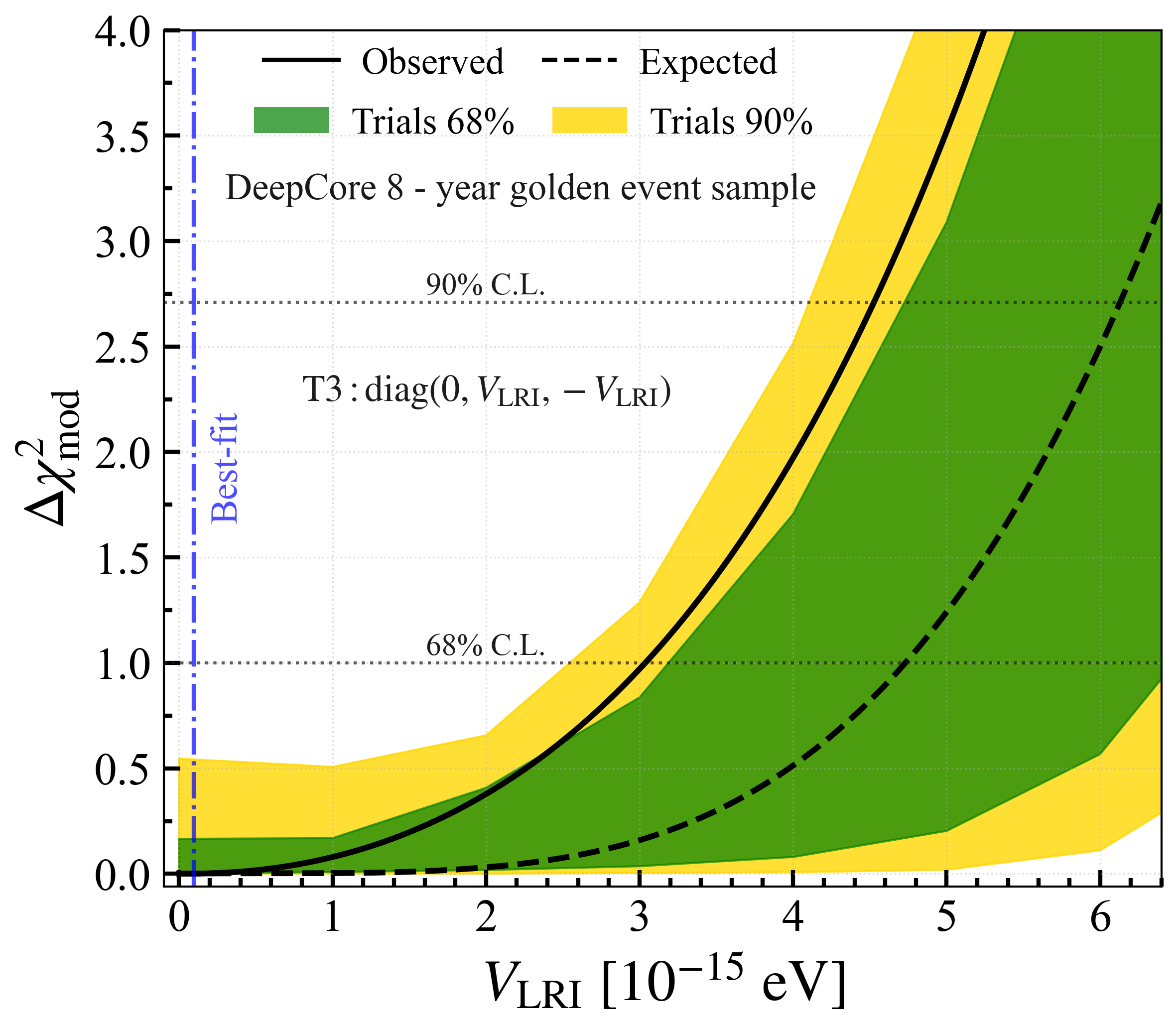}
 \caption{Constraints on the LRI potential sourced by the texture T3 of $U(1)^\prime$ symmetries using the 8-year high-purity $\nu_\mu$ CC dataset of IceCube DeepCore, assuming NO. The solid-black curve represents the observed $\Delta \chi^2_{\rm mod}$ as a function of $V_{\rm LRI}$, while the dashed-black curve shows the corresponding expected sensitivity obtained from MC. The vertical dashed-dotted blue line denote the best-fit value of $V_{\rm LRI}$. The green and yellow bands indicate the respective range of 68\% and 90\% from the distribution of $\Delta \chi^2_{\rm mod}$ which are obtained from fitting the 500 statistically fluctuated pseudo-experiments.
 	\label{fig:txt3}} 
\end{figure}

In this section, we describe the results obtained from the numerical analysis, which provide constraints on the LRI potential for the three textures T1, T2, and T3, considered one at a time. We perform a fit to the observed data using simulated MC events by varying the corresponding $V_{\rm LRI}$ parameter, along with all relevant nuisance parameters for each texture one at a time. The best-fit values of the LRI potentials are summarized in Table~\ref{tab:best-fit-table}. The respective best-fit values are also shown by the vertical dashed-dotted blue lines in Figs.~\ref{fig:txt1},~\ref {fig:txt2}, and~\ref{fig:txt3}, respectively. The best-fit values of nuisance parameters for each texture are listed in Table~\ref{tab:systematic_params} of the appendix~\ref{app:systematics}.

\begin{table}[ht]
	\centering
	\renewcommand{\arraystretch}{1.8}
	\begin{tabular}{ccc}
		\hline\hline
		Texture $~~$ & Best-fit Value (eV) $~~$ & Bound at 90\% C.L. (eV) \\
		\hline
		T1 & $1.59 \times 10^{-22}$ & $8.8 \times 10^{-15}$ \\
		T2 & $2.93 \times 10^{-16}$ & $9.1 \times 10^{-15}$ \\
		T3 & $9.71 \times 10^{-17}$ & $4.5 \times 10^{-15}$ \\
		\hline\hline
	\end{tabular}
	\caption{Best-fit values and 90\% C.L. upper bounds on the LRI potential $V_{\rm LRI}$ for the three textures considered in this analysis.}
	\label{tab:best-fit-table}
\end{table}

The test-statistic value from the data fitting for each texture is in agreement with the test-statistic observed for the fit with SI ($V_{\rm LRI} =  0$) case. This shows that the observed data prefer the SI scenario and does not show any evidence for the presence of LRI. Therefore, we place constraints on $V_{\rm LRI}$, for the three textures T1, T2, and T3, as shown in Figs.~\ref{fig:txt1},~\ref {fig:txt2}, and~\ref{fig:txt3}, respectively. To obtain these constraints, we perform a scan over $V_{\rm LRI}$ (one texture at a time), and compute the corresponding test statistic as,
\begin{equation}
	\Delta \chi^2_{\rm mod} = \chi^2_{\rm mod}(V_{\rm LRI}\ \text{fixed}) - \chi^2_{\rm mod}(V_{\rm LRI}\ \text{free}) ,
\end{equation}
 where $\chi^2_{\rm mod}$ is minimized over all the nuisance parameters.

The solid-black curves in Figs.~\ref{fig:txt1}--\ref{fig:txt3} represent the observed $\Delta \chi^2_{\rm mod}$ profile obtained by fitting the data. From these distributions, we extract the 90\% C.L. upper limits on $V_{\rm LRI}$ for each texture considering one degree of freedom. The dashed-black curves show the expected sensitivities derived from the simulated data. To estimate the expected sensitivity for a given texture, we construct a simulated dataset at the best-fit values of $V_{\rm LRI}$ and the nuisance parameters obtained from the data fitting (see Table~\ref{tab:systematic_params} in the appendix~\ref{app:systematics}), and compute the corresponding $\Delta \chi^2_{\rm mod}$ by profiling over $V_{\rm LRI}$. The green and yellow bands show the expected range of statistical variation in $\Delta \chi^2_{\rm mod}$ with 68\% and 90\% coverage, respectively. These bands are obtained by fitting the 500 pseudo-experiments produced with the best-fit values, where for each pseudo-experiment, the simulated events in each bin are statistically fluctuated. All upper limits at 90\% C.L. obtained from this work are of similar order ($10^{-15}$ eV) and are summarized in Table~\ref{tab:best-fit-table}. The constraint for the texture T3 is stronger in comparison to others, which is consistent with the observations in Figs.~\ref{fig:osc_prob}, \ref{fig:oscillogram}, and~\ref{fig:event_diff}.

\begin{figure}
	\centering
	\includegraphics[width=\linewidth]{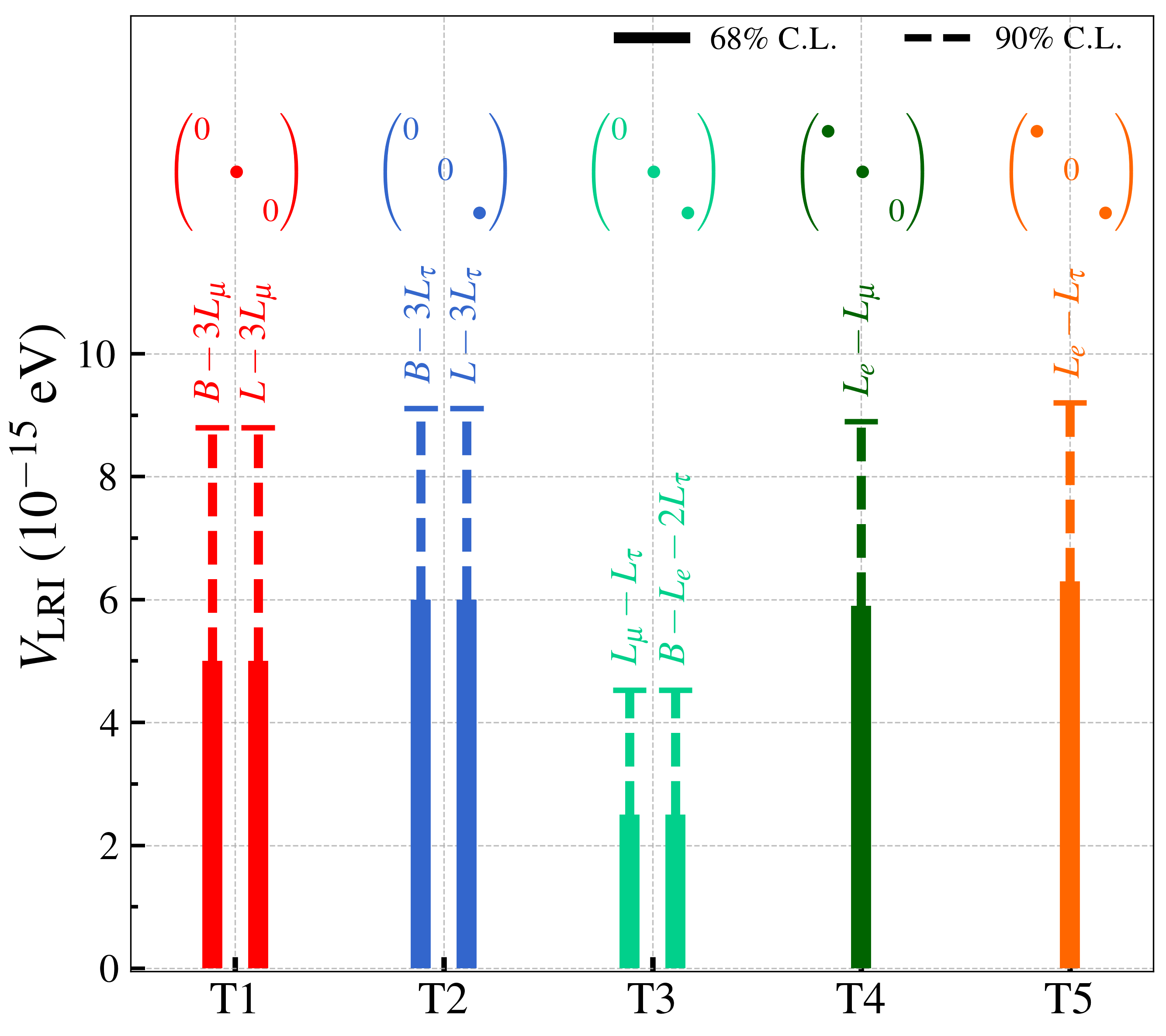}
	\caption{Constraints on the LRI potentials induced by
	different $U(1)^\prime$ symmetry textures assuming NO. The solid (dashed) vertical lines show the 68\% (90\%) C.L. constraints on $V_{\rm LRI}$ for all the textures considered one at a time. The $U(1)^\prime$ symmetries are grouped according to the texture of the induced LRI potential, as summarized in Table~\ref{tab:symm}, where symmetries leading to identical textures yield same constraints.}
	\label{fig:constraints_all}
\end{figure}

Figure~\ref{fig:constraints_all} summarizes the corresponding 68\% (solid lines) and 90\% (dashed lines) C.L. constraints on the LRI potentials for all the textures in a single panel. Symmetries that induce identical textures have same constraint, which is obtained by fitting the data only once for a given texture (see Figs.~\ref{fig:txt1},~\ref {fig:txt2}, and~\ref{fig:txt3} for T1, T2, and T3, respectively). For comparison, we also include the limits for the $L_e - L_\mu$ and $L_e - L_\tau$ symmetries from our previous study~\cite{Garg:2026gwx}, highlighting the relative sensitivities for different scenarios.

The 90\% C.L. bounds on the LRI potentials for each texture are also translated into constraints on the effective coupling strength and mass of ultra-light mediator, as shown in Fig.~\ref{fig:g_m_bound}. The limits from other experiments are also shown in Fig.~\ref{fig:g_m_bound} for comparison. Overall, our results provide some of the most stringent constraints using neutrino oscillations to date on the flavor-dependent long-range neutrino interactions arising from a broad class of $U(1)^\prime$ symmetries, with a comprehensive treatment of systematic uncertainties.

\section{Summary and conclusions}
\label{sec:conclusion}

In this work, we perform a comprehensive search for the flavor-dependent long-range neutrino interactions arising from a variety of anomaly-free $U(1)^\prime$ gauge symmetries mediated by ultra-light vector gauge bosons. Extending beyond the widely studied $L_e - L_\mu$ and $L_e - L_\tau$ scenarios, we investigate a broader class of viable $U(1)^\prime$ symmetries that generate long-range matter potentials sourced not only by electrons, but also by protons and neutrons distributed across astrophysical environments. Owing to the enormous number of matter constituents in these sources, even extremely feeble LRI couplings can induce sizable effective potentials that modify the propagation of neutrinos through Earth. Such effects can lead to observable distortions in neutrino oscillations and can therefore be probed using the available data sample of the terrestrial neutrino experiments.

Atmospheric neutrinos provide a unique avenue for probing such long-range interactions due to their wide energy range and long propagation distances through the Earth, where such new potentials can leave observable imprints. To probe these new interactions, we present a detailed analysis of the publicly available 8-year high-purity $\nu_\mu$ charged-current atmospheric neutrino data collected by IceCube DeepCore.
 
Using IceCube DeepCore data, we search for the deviations induced by these new LRI potentials with respect to the standard three-flavor neutrino oscillations. We find no evidence for these new interactions in DeepCore data sample, and therefore, place stringent constraints of the order of $10^{-15}$ eV on the corresponding LRI potentials. We translate the LRI potential constraints into that on the plane of mass and coupling strength of new ultra-light gauge boson. For mediator masses below $10^{-18}$ eV, our results provide some of the strongest limits to date over a wide range of mediator masses and coupling strengths for a broad class of $U(1)^\prime$ gauge symmetries.

Our analysis demonstrates the strong potential of atmospheric neutrino experiment like IceCube DeepCore to probe the properties of new ultra-light gauge bosons. A large amount of atmospheric neutrino data that will be available from the currently running experiments such as IceCube Upgrade~\cite{Ishihara:2019aao,IceCube:2025chb} and KM3NeT/ORCA~\cite{KM3Net:2016zxf}, as well as the next-generation experiments such as DUNE~\cite{DUNE:2021cuw} and Hyper-Kamiokande~\cite{Hyper-Kamiokande:2018ofw}, would play a crucial role in exploring the existence of these new long-range interactions in the neutrino sector.

\subsection*{Acknowledgments}

We acknowledge support from the Department of Atomic Energy (DAE), Govt. of India. G.G. acknowledges support from the Department of Science and Technology (DST), Govt. of India (Sanction Order No. DST/INSPIRE Fellowship/2021/IF210663). S.K.A., J.K., and A.K. receive support from the Swarnajayanti Fellowship (Sanction Order No. DST/SJF/PSA-05/2019-20) provided by the Department of Science and Technology (DST), Govt. of India, and the Research Grant (Sanction Order No. SB/SJF/2020-21/21) provided by the Anusandhan National Research Foundation (ANRF), Govt. of India, under the Swarnajayanti Fellowship. G.G. would like to thank P. Swain for useful discussions. The numerical simulations are performed using the Dell PowerEdge R660 Server at the Institute of Physics, Bhubaneswar, India.\\

\textbf{Data availability.}--- The data used to obtain results presented in this work are available in Ref.~\cite{DVN/B4RITM_2025}.\\

\begin{appendix}	
	
	\section{$U(1)^\prime$ charges of fermions}
	\label{app:U1-charges}

	Table~\ref{tab:charges} summarizes the assignment of $U(1)^\prime$ charges for different symmetries. The charges are determined directly from the definition of the underlying structure of the symmetry. For a generic symmetry of the form
	\begin{equation}
		U(1)^\prime = x_eL_e + x_\mu L_\mu	+ x_\tau L_\tau	+y B,
	\end{equation}
	the leptonic charges follow from the corresponding coefficients of the lepton numbers ($L_e, L_\mu,$ and $L_\tau$), while quarks carry charges equal to the product of coefficient $y$ and their baryon number (B). Accordingly, the coefficients $a_e$, $a_u$, and $a_d$ denote the $U(1)^\prime$ charges of the electron, up quark, and down quark, respectively, whereas $b_e$, $b_\mu$, and $b_\tau$ correspond to that of three neutrino flavors $\nu_e$, $\nu_\mu$, and $\nu_\tau$, respectively.
	
	Symmetries consisting of Lepton numbers only, are sourced by only electrons as muons and taus are not present in ordinary matter. On the other hand, symmetries involving baryon number receive the contributions from electrons, protons, and neutrons distributed in the Universe. Since nucleons are composite states of quarks, their effective $U(1)^\prime$ charges are obtained from the constituent quark charges as	
		\begin{align}
		a_p &=  2a_u + a_d,\\
		a_n &= a_u + 2a_d,
	\end{align}
	which corresponds to the quark compositions of the proton ($uud$) and neutron ($udd$), respectively. These effective nucleon charges enter directly in the evaluation of the induced LRI potential. In the last symmetry of Table~\ref{tab:charges}, $B_y = B_1 - yB_2 - (3-y)B_3$, where $B_1, B_2,$ and $B_3$ correspond to the baryon numbers of the first, second, and third generation of quarks, respectively, and $y$ denotes an arbitrary constant. The $U(1)^\prime$ charge assignments listed in Table~\ref{tab:charges} are used in Eq.~\ref{eq:total_pot} to compute $V_{\rm LRI}$ throughout this work.

\setcounter{table}{0}
\renewcommand{\thetable}{A\arabic{table}}
\begin{table}[tp!]
	\centering
	\renewcommand{\arraystretch}{1.3}
	\begin{tabular}{|c|c|c|c|c|c|c|c|}
		\hline
		\mr{2}{*}{$U(1)^\prime$ symmetry} &  \mc{6}{c|}{$U(1)^\prime$ charge}
		\\
		&\mr{1}{*}{$a_{u}$} & \mr{1}{*}{$a_{d}$} & \mr{1}{*}{$a_{e}$} & \mr{1}{*}{$b_{e}$} & \mr{1}{*}{$b_{\mu}$} & \mr{1}{*}{$b_{\tau}$} 
		\\
		\hline
		\mr{1}{*}{$B-3L_{\mu}$}& \mr{1}{*}{$\frac{1}{3}$}&\mr{1}{*}{$\frac{1}{3}$} & \mr{1}{*}{0} & \mr{1}{*} {0} & \mr{1}{*}{$-3$} & 0 
		\\
		\hline
		\mr{1}{*}{$L-3L_{\mu}$}& \mr{1}{*}{0}&\mr{1}{*}{0} & \mr{1}{*}{1} & \mr{1}{*} {1} & \mr{1}{*}{$-2$} & \mr{1}{*}{1}
		\\
		\hline
		\mr{1}{*}{$B-3L_{\tau}$}& \mr{1}{*}{$\frac{1}{3}$}&\mr{1}{*}{$\frac{1}{3}$} & \mr{1}{*}{0} & \mr{1}{*} {0} & \mr{1}{*}{0} & \mr{1}{*}{$-3$}
		\\
		\hline
		\mr{1}{*}{$L-3L_{\tau}$}& \mr{1}{*}{0}&\mr{1}{*}{0} & \mr{1}{*}{1} & \mr{1}{*} {1} & \mr{1}{*}{1} & \mr{1}{*}{$-2$}
		\\
		\hline
		\mr{1}{*}{$L_{\mu}-L_{\tau}$}& \mr{1}{*}{0}&\mr{1}{*}{0} & \mr{1}{*}{0} & \mr{1}{*} {0} & \mr{1}{*}{1} & \mr{1}{*}{$-1$} 
		\\
		\hline					
		\mr{1}{*}{$B-L_{e}-2L_{\tau}$}& \mr{1}{*}{$\frac{1}{3}$}&\mr{1}{*}{$\frac{1}{3}$} & \mr{1}{*}{$-1$} & \mr{1}{*} {$-1$} & \mr{1}{*} {0} & \mr{1}{*} {$-2$}
		\\
		\hline
		\mr{1}{*}{$L_{e}-L_{\mu}$}& \mr{1}{*}{0}&\mr{1}{*}{0} & \mr{1}{*}{1} & \mr{1}{*} {1} & \mr{1}{*}{$-1$} & \mr{1}{*}{0} 
		\\
		\hline
		\mr{1}{*}{$L_{e}-L_{\tau}$}& \mr{1}{*}{0}&\mr{1}{*}{0} & \mr{1}{*}{1} & \mr{1}{*} {1} & \mr{1}{*}{0} & \mr{1}{*}{$-1$} 
		\\
		\hline
		\mr{1}{*}{$B-3L_{e}$}& \mr{1}{*}{$\frac{1}{3}$}&\mr{1}{*}{$\frac{1}{3}$} & \mr{1}{*}{$-3$} & \mr{1}{*} {$-3$} & \mr{1}{*}{0} & \mr{1}{*}{0} 
		\\
		\hline
		\mr{1}{*}{$L-3L_{e}$}& \mr{1}{*}{0}&\mr{1}{*}{0} & \mr{1}{*}{$-2$} & \mr{1}{*} {$-2$} & \mr{1}{*}{1} & \mr{1}{*}{1} 
		\\
		\hline
		\mr{1}{*}{$B-\frac{3}{2}(L_{\mu}+L_{\tau})$}& \mr{1}{*}{$\frac{1}{3}$}&\mr{1}{*}{$\frac{1}{3}$} & \mr{1}{*}{0} & \mr{1}{*} {0} & \mr{1}{*}{$-\frac{3}{2}$} & \mr{1}{*}{$-\frac{3}{2}$} 
		\\
		\hline
		\mr{1}{*}{$L_{e}-\frac{1}{2}(L_{\mu}+L_{\tau})$}& \mr{1}{*}{0}&\mr{1}{*}{0} & \mr{1}{*}{1} & \mr{1}{*} {1} & \mr{1}{*}{$-\frac{1}{2}$} & \mr{1}{*}{$-\frac{1}{2}$} 
		\\
		\hline
		\mr{1}{*}{$L_{e}+2L_{\mu}+2L_{\tau}$}& \mr{1}{*}{0}&\mr{1}{*}{0} & \mr{1}{*}{1} & \mr{1}{*} {1} & \mr{1}{*}{2} & \mr{1}{*}{2} 
		\\
		\hline
		\mr{1}{*}{$B_{y}+L_{\mu}+L_{\tau}$}& \mr{1}{*}{$\frac{1}{3}$}&\mr{1}{*}{$\frac{1}{3}$} & \mr{1}{*}{0} & \mr{1}{*} {0} & \mr{1}{*}{1} & \mr{1}{*}{1} 
		\\
		\hline
	\end{tabular}
	\caption{The $U(1)^{\prime}$ charges of the fermions for the different symmetries. Charges $a_u$, $a_d$, and $a_e$ correspond to the up quark, down quark, and electron, respectively, whereas $b_e$, $b_\mu$, and $b_\tau$ denote that of $\nu_e$, $\nu_\mu$, and $\nu_\tau$, respectively. The charge for proton is $a_p = 2 a_u + a_d$ and that for neutrons is $a_n = 2 a_d + a_u$.}
	\label{tab:charges}
\end{table}
	
	\section{Neutrino oscillograms with LRI}
	\label{app:Oscillograms}
	\setcounter{figure}{0}
	\renewcommand{\thefigure}{B\arabic{figure}}
	\begin{figure}
		\centering
		\includegraphics[width=\linewidth]{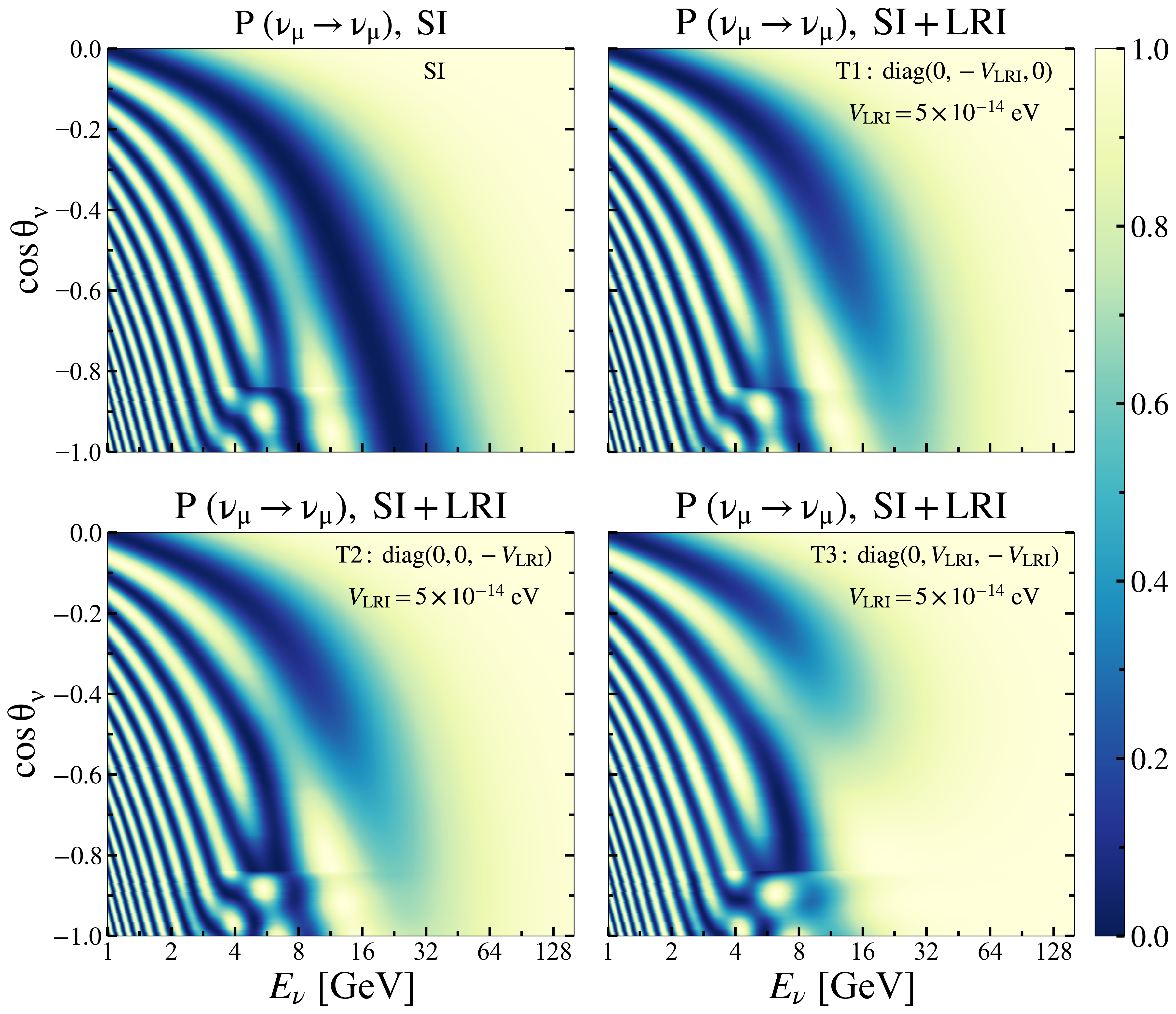}
		\caption{$P(\nu_\mu \rightarrow \nu_\mu)$ oscillograms in the $(E_\nu, \cos\theta_\nu)$ plane. The top-left panel corresponds to the SI case, while the top-right, bottom-left, and bottom-right panels correspond to the SI + LRI ($V_{\mathrm{LRI}} = 5 \times 10^{-14},\mathrm{eV}$) scenarios for different LRI textures, T1, T2, and T3, respectively. We assume NO with $\theta_{23} = 45.57^\circ$ and $\Delta m^2_{31} = 2.48 \times 10^{-3}~\mathrm{eV}^2$.}
		\label{fig:all_osc} 
	\end{figure}
	
	Figure~\ref{fig:all_osc} shows the probability oscillograms for P($\nu_\mu \rightarrow \nu_\mu$) as a function of neutrino energy and cosine of zenith angle. The top-left panel corresponds to the SI scenario without LRI, while the top-right, bottom-left, and bottom-right panels show probabilities for the three representative LRI textures T1, T2, and T3, respectively, with $V_{\rm LRI} = 5 \times 10^{-14} ~\text{eV}$. We can observe that the effect of LRI is more prominent in the region of oscillation valley corresponding to longer baselines and intermediate energies for all the textures. Among them, the T3 texture exhibits the largest deviation, consistent with its stronger impact on the $\mu$-$\tau$ sector.
	
	\section{Systematic Uncertainties and their Best-fit Values}
	\label{app:systematics}
	
	\setcounter{figure}{0}
	\renewcommand{\thefigure}{C\arabic{figure}}
	\begin{figure}
		\centering
		\includegraphics[width=\linewidth]{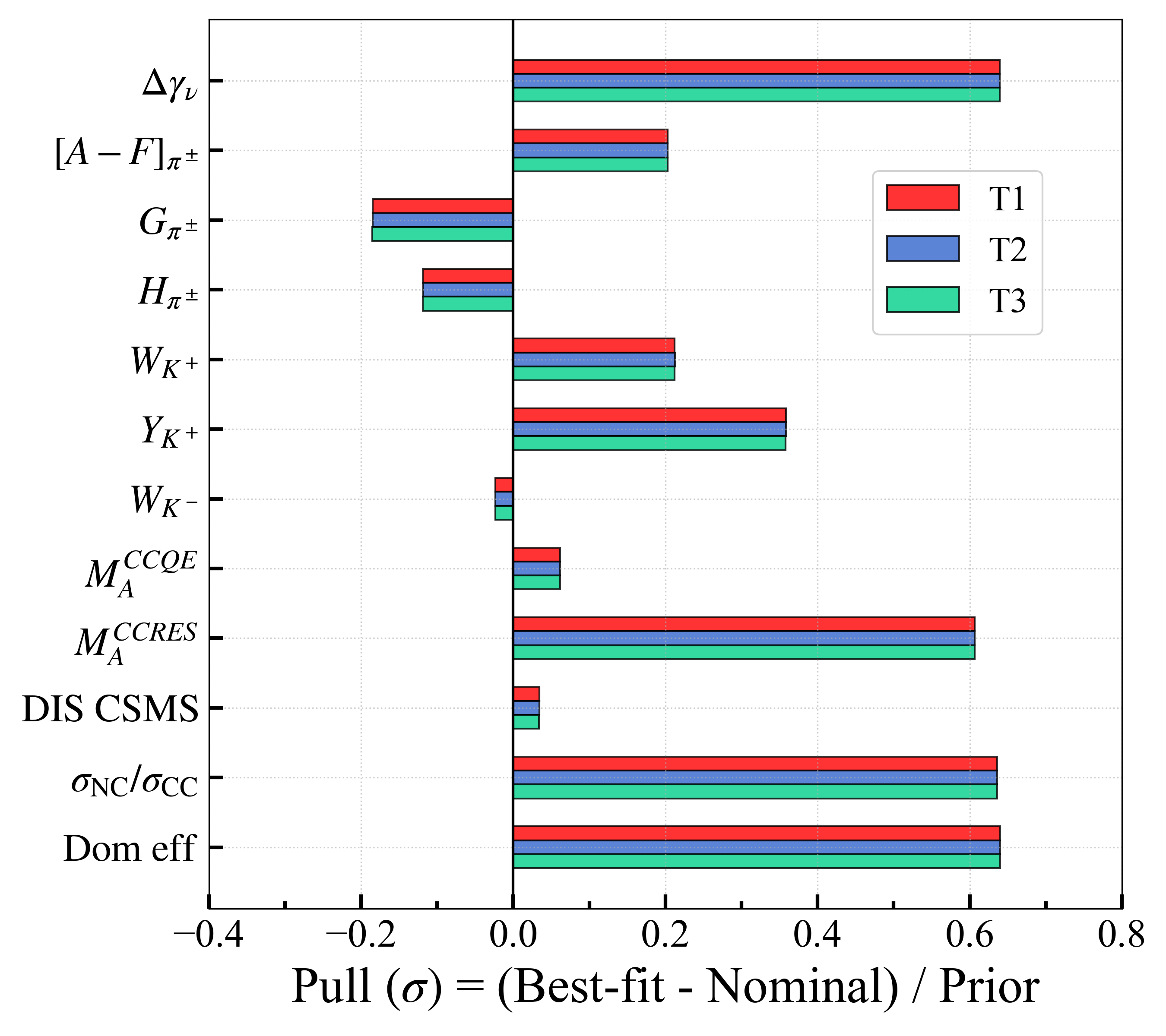}
		\caption{Pulls of the systematic parameters (if prior is available) obtained from the data fitting for the three representative LRI textures considered one at a time.}
		\label{fig:pull}
	\end{figure}
	
	The nuisance parameters corresponding to systematic uncertainties incorporated in this analysis are summarized in Tab.~\ref{tab:systematic_params}. The best-fit values obtained from the data fitting for the different LRI textures T1, T2, and T3, considered one at a time, are given in columns two, three, and four, respectively. The column five represents the nominal values of the parameters. The column six lists the Gaussian priors for some of the systematic parameters, whereas parameters without prior are left unconstrained during the minimization. The last column shows the allowed ranges for each parameter used in the minimization. Figure~\ref{fig:pull} shows the deviation of the best-fit values from their nominal values, expressed in terms of their priors. We can observe that all the parameters shown in Fig.~\ref{fig:pull} exhibit deviations of less than 1$\sigma$ relative to their respective prior. The pull of parameters are similar for the cases of all the three textures.
	
	
	\onecolumngrid
	
	\setcounter{table}{0}
	\renewcommand{\thetable}{C\arabic{table}}
	\begin{table}[h!]
		\centering
		\renewcommand{\arraystretch}{1.3}
		\begin{tabular}{l@{\hskip 15pt}c@{\hskip 15pt}c@{\hskip 15pt}c@{\hskip 15pt}c@{\hskip 15pt}c@{\hskip 15pt}c}
			\hline
			\hline
			\textbf{Parameter} & \textbf{Best-fit} (T1) & \textbf{Best-fit} (T2) & \textbf{Best-fit} (T3) &\textbf{Nominal value} & \textbf{Prior} & \textbf{Range} \\
			\hline
			\multicolumn{6}{@{}l}{\textbf{Detector:}} \\
			DOM efficiency		     & 1.064 	& 1.064 	&1.064	& 1.0 		& $\pm\,$0.1 & [0.8, 1.2] \\
			Ice absorption           & 0.974 	& 0.974 	& 0.974	& 1.0 		& -&[0.9, 1.1] \\
			Ice scattering           & 0.988 	& 0.988 	& 0.988	& 1.05 		& -&[0.95, 1.15] \\
			Relative eff. $p_0$      & $-\,0.269$ 	&      $-\,0.269$ 	& $-\,0.269$ & 0.10 		& -&[$-\,0.2$, 0.6] \\
			Relative eff. $p_1$      & $-\,0.043$ & $-\,0.043$ 	&  $-\,0.043$ & $-\,0.05$ 	& -&[$-\,0.2$, 0.2] \\
			\hline
			\multicolumn{6}{@{}l}{\textbf{Atmospheric flux:}} \\
			$\Delta \gamma_{\nu}$     & 0.064 & 0.064 & 0.064 & 0.0 & $\pm\,$0.1 & [$-\,0.5$, 0.5] \\
			$\Delta \pi^\pm \text{ yields [A-F]}$ & 0.061 & 0.061 & 0.061 & 0.0 & $\pm\,$0.3 & [$-\,1.5$, 1.5] \\
			$\Delta \pi^\pm \text{ yields G}$& $-\,0.055$ &  $-\,0.055$ & $-\,0.056$  & 0.0 & $\pm\,$0.3 & [$-\,1.5$, 1.5] \\
			$\Delta \pi^\pm \text{ yields H}$& $-\,0.018$ &  $-\,0.018$ &$-\,0.018$ & 0.0 & $\pm\,$0.15 & [$-\,0.75$, 0.75] \\
			$\Delta K^+ \text{ yields W}$ & 0.085 & 0.085 & 0.085 & 0.0 &   $\pm\,$0.4 & [$-\,2.0$, 2.0]\\
			$\Delta K^+ \text{ yields Y}$ & 0.107 & 0.108 & 0.107 & 0.0 &  $\pm\,$0.3 & [$-\,1.5$, 1.5]\\
			$\Delta K^- \text{ yields W}$ & $-\,0.009$ &  $-\,0.009$ & $-\,0.009$ &  0.0 & $\pm\,$0.4 & [$-\,2.0$, 2.0]\\
			\hline
			\multicolumn{6}{@{}l}{\textbf{Cross-section:}} \\
			$M_A^{\text{CCQE}}$ (in $\sigma$)  & 0.062 & 0.062 & 0.062 & 0.0 & $\pm\,$1.0 & [$-2.0$, 2.0]\\
			$M_A^{\text{CCRES}}$ (in $\sigma$)& 0.607 & 0.606 & 0.606 &  0.0 & $\pm\,$1.0 & [$-2.0$, 2.0]\\
			DIS CSMS                & 0.034 & 0.034 & 0.034 & 0.0 &  $\pm\,$1.0 & [$-3.0$, 3.0]\\
			$\sigma_{\rm NC}/\sigma_{\rm CC} $ & 1.127 & 1.127 & 1.127 &  1.0 & $\pm\,$0.2 & [0.5, 1.5]\\
			\hline
			\multicolumn{6}{@{}l}{\textbf{Normalization:}} \\
			$A_{\text{eff}}$ scale   & 0.824 & 0.824 & 0.824 & 1.0&  -&[0.6, 1.4] \\
			\hline
			\multicolumn{6}{@{}l}{\textbf{Atmospheric muons:}} \\
			Atm. $\mu$ scale         & 1.365 & 1.365 & 1.365 & 1.0&  -&[0.7, 1.5] \\
			\hline
			\multicolumn{6}{@{}l}{\textbf{Oscillations:}} \\
			$\theta_{23}$           & 45.372$^\circ$ & 45.326 &  45.341$^\circ$ & 45.573$^\circ$ & -&[38$^\circ$, 52$^\circ$] \\
			$\Delta m^2_{31}$       & 0.002489 eV$^2$ & 0.002489 eV$^2$ & 0.002489 eV$^2$ & 0.002484 eV$^2$ & -&[0.002, 0.003] eV$^2$ \\
			\hline
			\hline
		\end{tabular}
		\caption{A summary of the systematic uncertainty parameters incorporated in this analysis as nuisance parameters, along with their nominal values (fifth column), available $1\sigma$ priors (sixth column), and allowed fitting ranges (seventh column). The second, third, and fourth columns present the best-fit values obtained under the hypotheses of T1, T2, and T3 textures, respectively, considered one at a time assuming NO.}
		\label{tab:systematic_params}
	\end{table}
	\clearpage
	\twocolumngrid
	
	\section{Data-MC agreement}
	\label{app:data-mc}

	Figure~\ref{fig:chi2_dist} shows the observed $\chi^2$ values for the three textures T1, T2, and T3, considered in this work along with their expected $\chi^2$ distributions obtained from pseudo-experiments incorporating statistical fluctuations. The $p$-values representing the goodness-of-fit are calculated as the fraction of pseudo experiments having $\chi^2_{\rm mod}$ greater than the observed $\chi^2_{\rm mod}$. All three textures show p-values of around $0.25$.
	
	Figures~\ref{fig:event_energy} and \ref{fig:event_zenith} show the one-dimensional projections of the data-MC comparison for the $E_{\rm reco}$ and $\cos \theta_{\rm reco}$, respectively. The distributions of observed data are shown by the black dots with error bars.
	The best-fit distributions for T1, T2, and T3 textures are shown by the red, blue, and cyan colors, respectively. Additionally, the distributions of expected events for a representative choice of $V_{\rm LRI} = 5\times10^{-14}$ eV for all the three textures are also shown to demonstrate the signal regions in $E_{\rm reco}$ and $\cos \theta_{\rm reco}$.
	
	\setcounter{figure}{0}
	\renewcommand{\thefigure}{D\arabic{figure}}
	\begin{figure}[h]
		\centering
		\includegraphics[width=\linewidth]{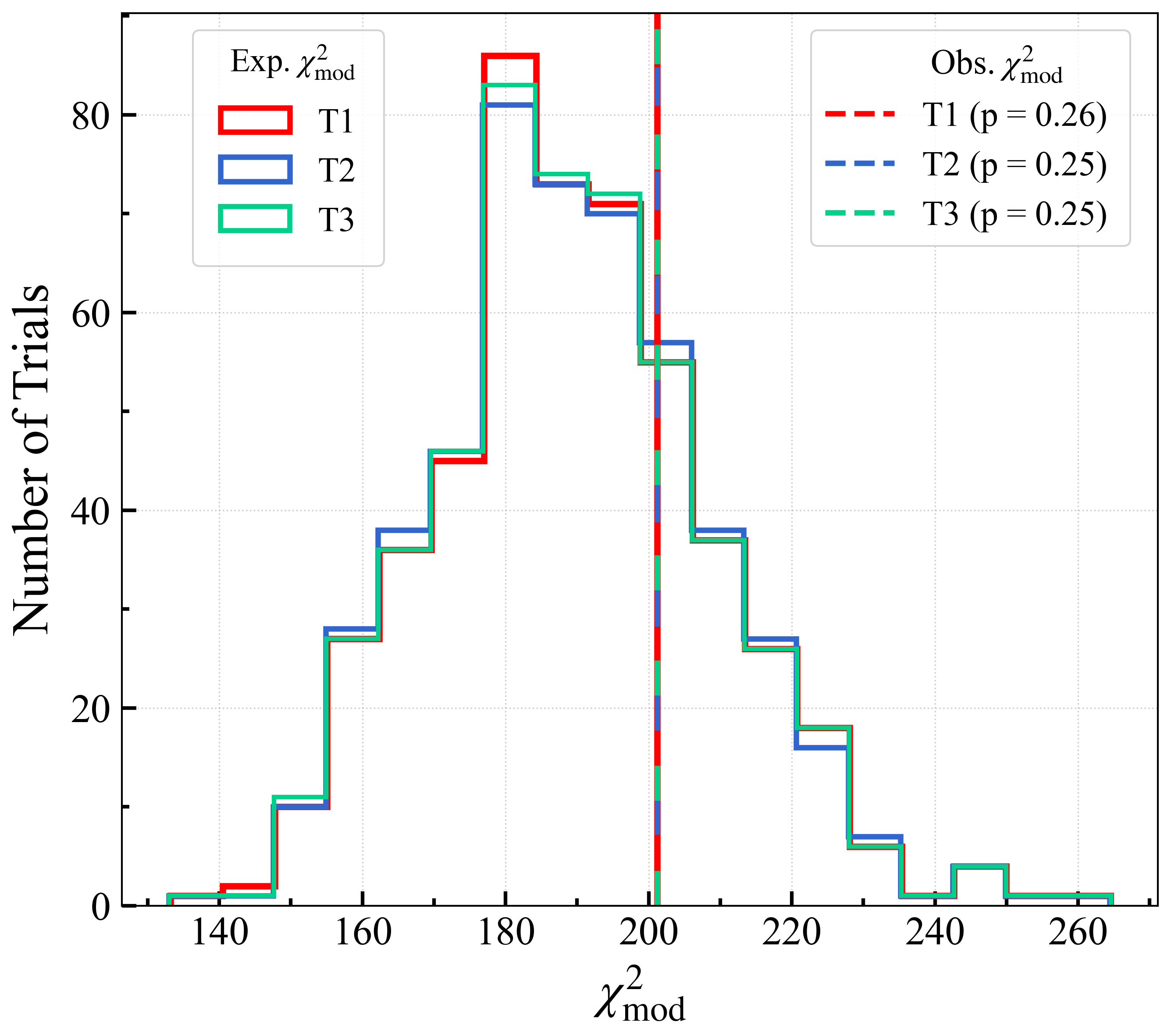}
		\caption{Comparison between the observed $\chi^2_{\rm mod}$ (shown as vertical lines) and the expected distributions derived from statistically fluctuated 500 pseudo-experiments. Each distribution corresponds to one of the three LRI textures fitted separately. The resulting $p$-values are around 0.25 for all the three textures.}
		\label{fig:chi2_dist}
	\end{figure}
	
	\begin{figure}[h]
		\centering
		\includegraphics[width=\linewidth]{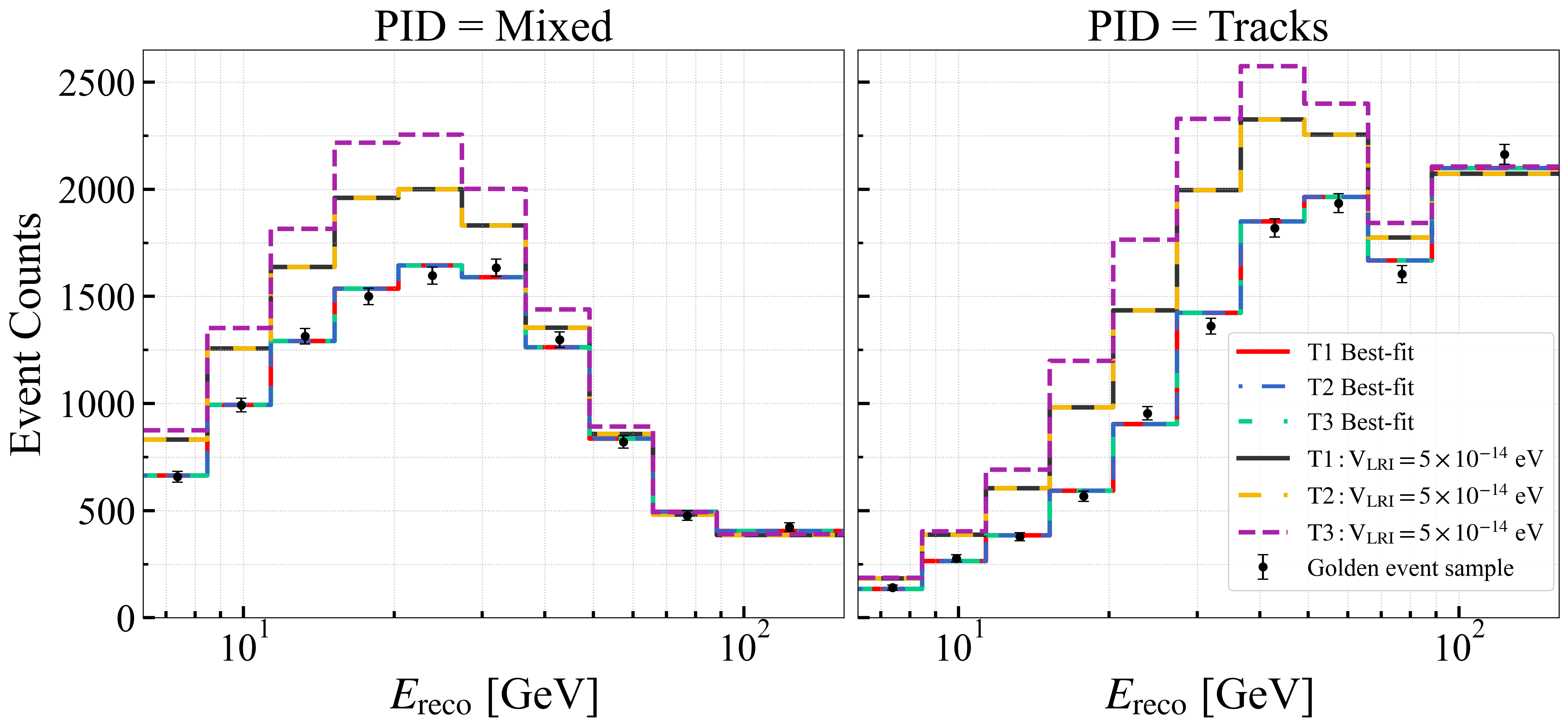}
		\caption{Distributions of $E_{\rm reco}$ comparing the observed 8-year DeepCore data with the best-fit MC predictions for the three LRI textures considered one at a time. For comparison, distributions for the representative choice of $V_{\rm LRI}=5\times10^{-14}$ eV are also included for the three textures. The left and right panels show the distributions for mixed and track-like events, respectively.}
		\label{fig:event_energy}
	\end{figure}
	
	\begin{figure}[h]
		\centering
		\includegraphics[width=\linewidth]{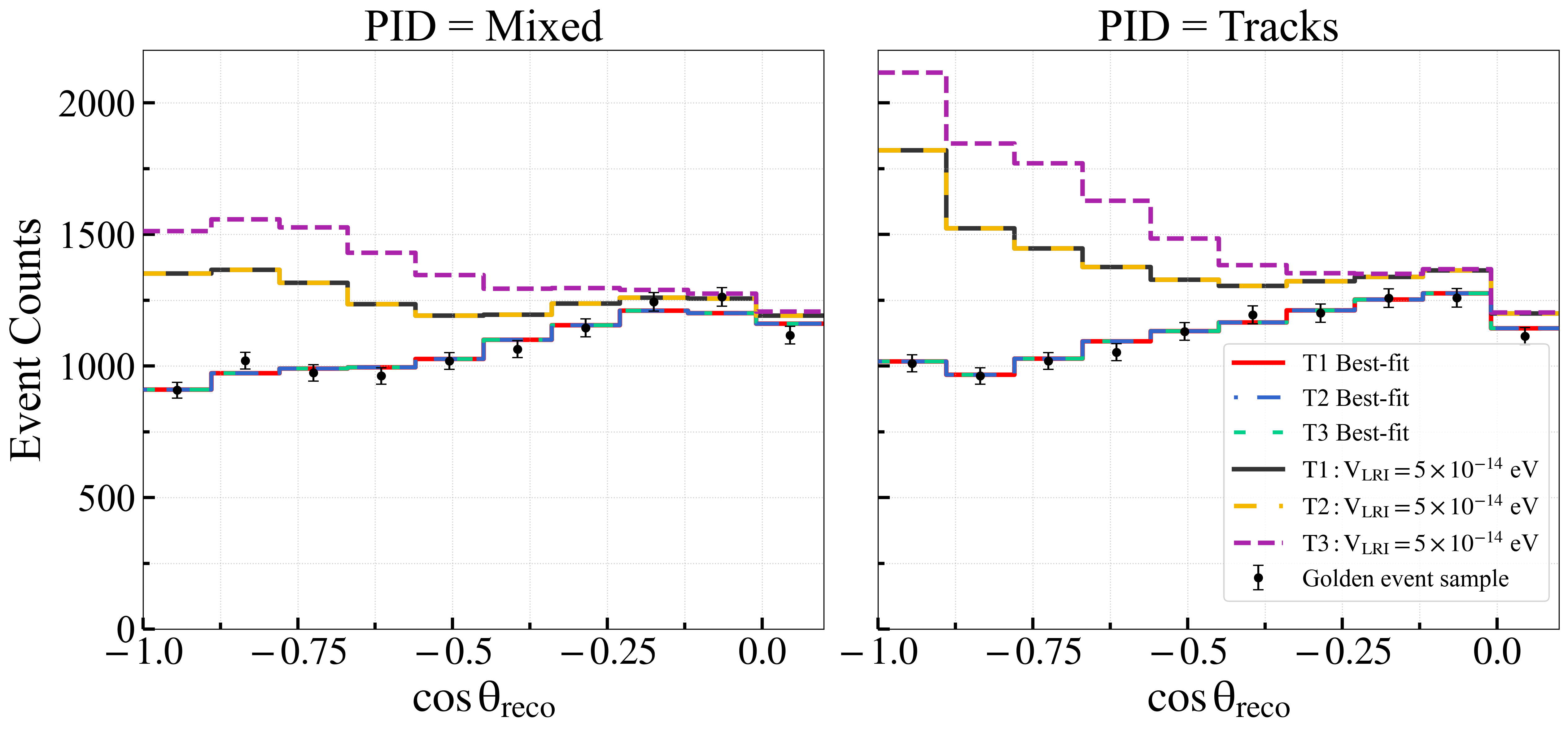}
		\caption{Distributions of $\cos \theta_{\rm reco}$ comparing the observed 8-year DeepCore data with the best-fit MC predictions for the three LRI textures considered one at a time. For comparison, distributions for the representative choice of $V_{\rm LRI}=5\times10^{-14}$ eV are also included for the three textures. The left and right panels show the distributions for mixed and track-like events, respectively.}
		\label{fig:event_zenith}
	\end{figure}

\end{appendix}
\newpage

\bibliography{refs}

\end{document}